\documentclass[11pt,a4paper]{article}
\pdfoutput=1 
\usepackage{jheppub}	
\usepackage{mathabx}
\usepackage{tikz}
\usetikzlibrary{positioning,arrows}
\usetikzlibrary{decorations.pathmorphing}
\usetikzlibrary{decorations.markings}
\usetikzlibrary{snakes}
\usetikzlibrary{matrix}

\usepackage{epsfig}
\usepackage{graphicx}
\usepackage[vcentermath,enableskew]{youngtab}

\usepackage[utf8]{inputenc}
\usepackage[english]{babel}
\usepackage{makeidx}
\usepackage{tikz}
\tikzset{every picture/.style={line width=0.75pt}} 
\tikzset{line/.style={thick, decorate, draw=black,}}
\usepackage{amsfonts}
\usepackage{enumerate}
\usepackage{mathrsfs}
\usepackage{tensor}
\usepackage[autostyle]{csquotes}
\usepackage{subfig}

\def\be{\begin{equation}}
\def\ee{\end{equation}}
\def\bea{\begin{eqnarray}}
\def\eea{\end{eqnarray}}
\newcommand{\nn}{\nonumber}

\newcommand{\ft}[2]{{\textstyle\frac{#1}{#2}}}
\def\ii{{\rm i}}

\def\apjl{\ref@jnl{ApJ}}

\usepackage{amsmath,amssymb}
\usepackage{latexsym}
\usepackage{graphicx}
\usepackage{slashed}

 \usepackage[vcentermath,enableskew]{youngtab}
\usepackage{epsfig}

\def\be{\begin{equation}}
\def\ee{\end{equation}}
\def\bea{\begin{eqnarray}}
\def\eea{\end{eqnarray}}

\def\cF{\mathcal{F}}

\title{ Post Newtonian emission of gravitational waves from binary systems: a gauge theory perspective}

\author[a]{Francesco Fucito}
\author[a]{and Jose Francisco Morales}\affiliation[a]{Sezione INFN  ``Roma Tor Vergata" and Dipartimento di Fisica, Università di Roma ``Tor Vergata", Via della ricerca
scientifica 1, 00133, Roma, Italy}

\abstract{Using the AGT correspondence and localization, we derive a combinatorial formula for the Post-Newtonian expansion of the
wave form describing the gravitational emission from  binary systems made of objects  of extremely different masses. The results
are written as a double instanton series describing the expansion of the gravitational wave at large distances and small velocities, and are tested against previous formulae  in the literature for Schwarschild and Kerr black holes at the 5th and 3rd Post Newtonian order respectively beyond the quadrupole approximation. Tidal effects show up in the wave form at the 5th PN order, providing a quantitative measure of sizes and reflectivity properties of the gravity solution.
}

\emailAdd{francesco.fucito,francisco.morales@roma2.infn.it}
\begin{document}
\tikzset{
line/.style={thick, decorate, draw=black,}
 }

\maketitle

 \section{Introduction}

 The detection of gravitational waves (GW) by the LIGO-Virgo collaboration \cite{LIGOScientific:2016aoc} is definitively one of  the most exciting achievements in contemporary physics.To extract the signal from the experimental data, templates of waveforms have been meticulously crafted
 through a sophisticated blend of theoretical  and numerical calculations simulating the three phases (inspiral, merging, ring down) of the coalescence of black holes (BH) and/or neutron stars. The matching against observations provides a strong test of general relativity and confirms the existence of black hole like objects in nature.
The upgrade of the LIGO-Virgo experiments and the coming of the ET experiment and the space based LISA will open a new era of precision gravitational wave (GW) astronomy  allowing to test gravity beyond the realm of general relativity.

  A fundamental question is to prove the existence of BH's in nature. These mathematically beautiful  solutions of the Einstein equations exhibit, at the classical level, remarkable  uniqueness and stability properties. At the quantum level instead, a BH radiate and evaporates completely, burning all the information used to built it, in contradiction with the fundamental principle of unitarity in physics.   Alternatives to black holes exist, and go under the collective name of Exotic Compact Objects (ECO). Examples are gravastars, wormholes, boson stars and fuzzballs \cite{Mathur:2005zp,Cardoso:2016rao,Cardoso:2017cqb,Cardoso:2019rvt}. ECO's are described by smooth horizonless geometries and can be distinguished from BH's by their multipolar structure \cite{Bena:2020see,Bianchi:2020bxa,Bena:2020uup,Bianchi:2020miz,Mayerson:2020tpn,Bah:2021jno}, quasi normal modes (QNM's)  and their characteristic echo based GW emission \cite{Ikeda:2021uvc}.  GW signals coming from a binary system can be described in analytic terms in the ringdown and inspiral phases.  In
   \cite{Aminov:2020yma,Bonelli:2021uvf,Bianchi:2021mft,Bonelli:2022ten,Bianchi:2022wku,Consoli:2022eey,Bianchi:2023sfs,Aminov:2023jve}, novel techniques
  capitalizing on recent advances in quantum field theories have been exploited to study the QNM's characterising the ringdown phase of the GW emission. More  traditional  analytical approaches to the dynamics of gravitationally interacting binary systems \cite{Blanchet:2013haa,Damour:2016gwp,Damour:2017zjx,Bini:2023fiz} have been  successfully compared against gauge inspired scattering amplitude techniques \cite{Bjerrum-Bohr:2018xdl,Cheung:2018wkq,Bern:2019nnu,KoemansCollado:2019ggb,Bern:2019crd,Bjerrum-Bohr:2019kec,Bern:2020gjj,DiVecchia:2021ndb,DiVecchia:2021bdo,Cristofoli:2021vyo,Herrmann:2021tct,Brandhuber:2021eyq,Bjerrum-Bohr:2021vuf,Bjerrum-Bohr:2021din,Adamo:2022qci,Brandhuber:2023hhy,DiVecchia:2023frv,Brandhuber:2023hhl,DeAngelis:2023lvf}, effective field theories \cite{Kalin:2020mvi,Kalin:2020fhe,Liu:2021zxr,Bjerrum-Bohr:2022ows} in the description of the inspiral phase. Other techniques like worldline quantum field theories \cite{Mogull:2020sak,Jakobsen:2021smu,Jakobsen:2021lvp,Jakobsen:2023ndj,Jakobsen:2022psy},
 and  two dimensional conformal field theories \cite{Novaes:2014lha,CarneirodaCunha:2015hzd,CarneirodaCunha:2015qln,Amado:2017kao,Lencses:2017dgf,BarraganAmado:2018zpa,Novaes:2018fry,CarneirodaCunha:2019tia,Amado:2020zsr,BarraganAmado:2021uyw,Cavalcante:2021scq,daCunha:2021jkm}
     originally devised for different fields of research, were used for comparison with gravity computations.

     In this paper we deal with the study of the inspiral phase of binary systems. We work in the extreme mass ratio limit of the inspiral motion (EMRI) where the light particle orbits for a long time around the heavy one accumulating  thousands of radians bringing tidal heating  in the sensitivity band of a space based interferometer like LISA \cite{Hughes:2001jr}.  In this regime the GW emission is well approximated by the the Post-Newtonian (PN) limit which assumes that the distance between the two objects is large and the relative velocity small.  The theoretical studies of  gravitational systems in the PN limit have a long history, and are very well described in many excellent textbooks \cite{Straumann:1984xf,Maggiore:2007ulw,Maggiore:2018sht,PoissonWill:2014}. These results have been checked and further expanded using BH perturbation theory. In this framework,  the Einstein equations are expanded to linear order in the perturbations around the background  and written in terms of  an ordinary second order differential equation of the Heun type resembling a Schr\"odinger equation.
   A binary system can also be described along these lines introducing in the equations a source term. The PN results can now be checked against this independent computation. Early results appeared in \cite{Poisson:1993vp} and were later expanded in a series of papers \cite{Poisson:1993vp,Tagoshi:1993dm,Tagoshi:1994sm,Shibata:1994jx,Tanaka:1996lfd,Mino:1997bx,Fujita:2010xj,Fujita:2011zk,Fujita:2012cm}  to extremely high orders for binary system orbiting circularly around Schwarzschild and Kerr BH's.

      We revisit the computation of the PN expansion of the waveform using the  CFT/gravity correspondence developed in \cite{Bonelli:2021uvf,Bonelli:2022ten,Consoli:2022eey}.   This line of research is founded on the so called AGT correspondence \cite{Alday:2009aq} which relates the results of localization for the partition functions of a $N=2$ supersymmetric quiver gauge theories \cite{Nekrasov:2002qd,Flume:2002az,Bruzzo:2002xf,Nekrasov:2003rj} with certain correlators of the Liouville CFT.  The crucial point is that some correlators in the two-dimensional CFT satisfy a second order differential equation \cite{Belavin:1984vu} which can be mapped to those of the Heun type which describe the wave form in the BH perturbation theory. As a result, the instanton gauge theory partition functions provide a combinatorial representation of the Heun function and its connection formulae \cite{Bonelli:2022ten,Consoli:2022eey}. Following these ideas we relate
      the PN description of the wave form in the near zone to the instanton partition function of a quiver gauge theory given as a double series in the couplings which are related to the PN expansion parameters. As a result the PN expansion is given as a series with coefficients given by rational functions of the orbital quantum number $\ell$. All logs and transcendental numbers that typically appear in the PN expansions are summed up into exponentials and Gamma functions. The instanton series is known to be convergent \cite{Its:2014lga,Arnaudo:2022ivo} and as a consequence so is the PN expansion.

      Tidal effects are expected to show up in the gravitational wave signal at the 5th PN order  ($v^{10}$) when the two independent solutions of the Heun equation start to mix \cite{Bildsten:1992my,Tanaka:1996lfd}. Their effects
      are proportional to the dynamical Love number defined in \cite{Charalambous:2021mea,Consoli:2022eey} and provide a quantitative measure of the compactness of the gravitational object. Along these lines we compute the static and dynamic Love numbers of Kerr-like compact objects. Recent results on this subject appeared recently in \cite{Maggio:2021uge}.

      This is the plan of the paper: In Section 2 we introduce the binary system and review the derivation of the wave form and luminosity
   of the gravitational wave signal. In section 3, we describe the CFT gravity dictionary adapting previous results \cite{Consoli:2022eey} to the needs of the present discussion. In Section 4 and 5 we  derive the PN expansion of the the wave form and its luminosity for Schwarschild and Kerr BH's at the 5th and 3rd PN order beyond the quadrupole approximation. In Section 6 we discuss the tidal interactions and point out the differences between BH's and ECO's.
  Appendices \ref{cfttools}, \ref{sourceterm}, \ref{aExpansion}, and \ref{mathrin} collect some background and technical material.

\section{ The binary system}

In this section we briefly review some background material about binary systems following \cite{Tagoshi:1993dm,Tagoshi:1994sm,Shibata:1994jx,Tanaka:1996lfd,Mino:1997bx,Fujita:2010xj,Fujita:2011zk,Fujita:2012cm} whose notations we follow.  We use units such that $c=G=1$ and adopt a mostly plus Lorentzian metric.

\subsection{Teukolsky equations}

The dynamics of the linear perturbations of the Kerr metric
\be
ds^2=\frac{\Delta_r(r)}{\Sigma}(dt-a \sin^2\theta d\phi)^2+\frac{\sin^2\theta}{\Sigma}\left[(r^2+a^2)d\phi-a\,dt
\right]^2+\frac{\Sigma}{\Delta_r(r)}dr^2+\Sigma d\theta^2
\ee
with $\Delta_r(r) = (r-r_+)(r-r_-)=r^2 -2 M r  +a^2,\, \Sigma=r^2+a^2\cos^2\theta$
can be decomposed into spin $s$ components, $\Psi_{s}$, obtained as projections
of the curvature tensors deviation from the Kerr metric. The Einstein equations for $\Psi_{s}$ can be separated into radial and angular components via the ansatz
\bea
\Psi_{s} &=&   \int d\omega \, \sum_{\ell,m} e^{{\rm i} m \phi-{\rm i}\omega t}  R_{\ell m}(r) {}_{s} S_{\ell m}(\chi)
 \label{teuk}
\eea
with  $\chi=\cos\theta$, and the functions $R_{\ell m}(r)$, $ {}_{s} S_{\ell m}(\chi)$ satisfying the Teukolsky equations \cite{Teukolsky:1972my}
 \bea
&&{d\over d\chi} \left[ \Delta_\chi (\chi ) {}_s S_{\ell m}'(\chi ) \right]   + \left(-\frac{(m+s \chi )^2}{1-\chi ^2}+s+A+ a^2 \chi ^2 \omega ^2-2 a s \chi  \omega\right) {}_s S_{\ell m}(\chi )=0 \label{teukang}\\
&&{1\over \Delta_r(r)^{s} }{d\over dr} \left[ \Delta_r(r)^{s+1} R_{\ell m}'(r) \right] { +}\left(\frac{K(r)^2{-}2{\rm i} s (r{-}M) K(r) }{\Delta_r(r)}{+}4{\rm i} s \omega r{-}\lambda_r  \right)R_{\ell m}(r)  =  {}_s T_{\ell m}(r) \nn
\eea
Here $M$ is the mass, $a=J/M$, the
angular momentum parameter, $A=(\ell-s)(\ell+s+1)+O(\omega a)$ is a separation constant and
\be
\Delta_\chi(\chi)=1-\chi^2 ,\quad
\lambda_r = A+a^2\omega^2-2am \omega,\quad    K(r)= (r^2+a^2) \omega-a m
\ee
Finally
 \begin{equation}
 \label{eqsource0}
{}_s T_{\ell m} (r) ={1\over \sqrt{2\pi}} \int d\chi\, d\phi \, dt  \, {}_{s}S_{\ell m} (\chi)\, e^{-im\phi+i\omega t}  \Sigma(r,\chi)\,   {}_s {\cal T}({\bf x})
\ee
%
%
%
is the Fourier transform of the spin $s$ stress-energy tensor ${}_s {\cal T}({\bf x}) $. We refer the reader to \cite{Teukolsky:1972my} for details. We are interested in the spin $s=-2$ mode associated to the $\psi_4$-perturbation.
The explicit form of $ {}_s T_{\ell m}$ has been derived in
\cite{Tagoshi:1993dm,Tagoshi:1994sm,Shibata:1994jx,Tanaka:1996lfd,Mino:1997bx,Fujita:2010xj,Fujita:2011zk,Fujita:2012cm}. For the reader's convenience, we collect in Appendix \ref{sourceterm}, the formulae relevant to the computations in this paper.

\subsection{The Green function}

The radial Teukolsky equation with a non-trivial source ${}_s T_{\ell m}(r)$ can be solved with the method of the Green function.
  For simplicity, we shall omit in the following the subscripts $\ell,m$. Given $\ell,m$, we denote by $T$ the source term, by $R_T$ the solution of the inhomogenous equation with incoming boundary conditions at the horizon and by $R_{\rm in,up}$ the solutions of the homogenous equation,  satisfying  incoming boundary conditions at the horizon and outgoing ones at infinity respectively.  The Green function will be written in terms of  $R_{\rm in}$ and $R_{\rm up}$.
 More explicitly, we write
\be
R_{T}(r)=\int_{r_+}^\infty G(r,r') \,  {}_{-2} T(r')   \Delta_r(r')^{-2} dr'. \label{rinhom}
\ee
with $G(r,r')$ the Green function with incoming boundary conditions at the horizon. Explicitly
\be
G(r,r')={1 \over  W}  \times  \left\{
\begin{array}{ccc}
	R_{ \rm in} (r)  R_{\rm up} (r')   & ~~~~~~~~~~~~~~~~~~~~  & r< r'   \\
	R_{\rm in} (r')  R_{ \rm up} (r)   & ~~~~~~~~~~~~~~~~~~~~  & r> r'   \\
\end{array}
\right.
\ee
with  $R_{\rm in/up}$  satisfying the boundary conditions
\bea
R_{\rm in} (r)  & \approx &
\left\{
\begin{array}{lll}
	D_{\rm in}  \Delta_r(r)^2 e^{-{\rm i}  k r_* } &   & \quad
	r\to r_+   \\
	B_{\rm up}   r^{3}  e^{ {\rm i} \omega r_*}   + B_{\rm in}    r^{-1}  e^{- {\rm i} \omega r_*}&~~~~~~~   &\quad r \to \infty   \\
\end{array}
\right. \nn\\
R_{\rm up} (r)  & \approx &
\left\{
\begin{array}{lll}
	\widetilde{D}_{\rm up} \,   e^{{\rm i}  k  r_* }  + \widetilde{D}_{\rm in} \,    \Delta_r(r)^2 e^{-{\rm i}  k  r_* }    & ~~~~~~~~~~~~~~~~~~~~~~~~~~~~~  & r \to r_+\\
	\widetilde{B}_{\rm up}  r^{3}  e^{ {\rm i} \omega r_*}   & ~~~~~~~~~~~~  &r \to \infty   \\
\end{array}
\right. \label{bcpsi}
\eea
where $k=\omega-ma/2Mr_+$,  $r_*= r+ \frac{2 M r_+}{r_+-r_-}  \log\frac{r-r_+}{2M}   -\frac{2 M r_-}{r_+-r_-}  \log\frac{r-r_-}{2M} $ is the tortoise coordinate and $W$ the Wronskian
\be
W=\left[ R_{\rm in} (r)  R_{\rm up}' (r) - R_{\rm in}' (r)  R_{\rm up} (r)  \right]/ \Delta_r(r) =2{\rm i} \omega \, B_{\rm in} \widetilde B_{\rm up} \label{ww}
\ee
Here we use the fact  that $W$ is a constant, as follows from the wave equation, so it has been evaluated  at $r\to \infty$.
Putting altogether one finds
\be
R_{T}(r) \underset{r\to \infty}{\approx}   r^3   e^{ {\rm i} \omega r_*}  Z_{\ell m}\label{rt00}
\ee
with
\be
Z_{\ell m} =  \int_{r_+}^\infty   { R_{ \rm in} (r')   \over 2 {\rm i}  \omega B_{\rm in}
}   {}_{-2} T(r')  \Delta_r(r')^{-2}  dr' \label{rt0}
\ee

\subsection{Circular orbits}

The  stress energy tensor produced by the  particle motion is given by
\be
T^{\mu\nu} = \frac{1}{\sqrt{-g}}\int d\tau  \dot{x}^\mu \dot{x}^\nu   \delta^{(4)} \left[  x-x_0(\tau) \right] \label{tt}
\ee
with $x_0(\tau)$ the geodesic trajectory.  In the Hamiltonian formalism, the dynamics of a particle moving along a geodetics of the Kerr geometry is governed by the  separable Hamiltonian
\be
{\mathcal H}=\ft12 (g^{\mu\nu} P_\mu P_\nu +\mu^2) =  {\mathcal H}_r+{\mathcal H}_\theta    =0
\ee
with $\mu$ the particle mass,
\begin{equation}
\label{hmltnn}
\begin{aligned}
2\Sigma \mathcal{H}_r &=P_r^2 {\Delta_r}-\frac{\left( P_t (a^2+r^2)+a {P_\phi}  \right)^2 }{{\Delta_r}}   +(a P_t+P_\phi)^2   -  \mu^2 r^2 -C =0 \\
2 \Sigma \mathcal{H}_\theta &=   P_\theta^2  +P_\phi^2 \cot^2\theta -P_t^2 a^2 \cos\theta^2 - \mu^2 a^2 \cos\theta^2 +C =0
\end{aligned}
\end{equation}
and $C$ a separation constant.  We are interested in circular geodesics along the equatorial plane $\theta=\pi/2$.
 They are defined by the solutions of
\be
 P_\theta(\theta)=P_r(r)=\partial_r P_r(r)=0
\label{ppp}\ee
 with $P_\theta(\theta)$, $P_r(r)$ obtained by solving (\ref{hmltnn}).
It is convenient to introduce the parameters $v$, $\mathfrak{q}$ to parametrize the radius $r_0$ of the orbit and the angular parameter $a$ via the identifications
\be
r_0= M v^{-2} \quad , \quad a=\mathfrak{q}  M
\ee
In terms of these variables the solutions of  (\ref{hmltnn}) and  (\ref{ppp})
are
\be
P_t=- \mu {1-2v^2+\mathfrak{q} v^3\over \sqrt{1-3v^2+2\mathfrak{q}v^3}} \quad ,\quad  P_\phi= {\mu M \over  v }  {(1-2\mathfrak{q}v^3+\mathfrak{q}^2 v^4) \over
\sqrt{1-3v^2+2\mathfrak{q}v^3}}   \quad , \quad C=0 \label{ptpp}
\ee
The geodesic orbit is determined by integrating the velocities
 \be
 \dot{x}^\mu={d x^\mu \over d\tau}={\partial {\cal H}\over \partial P_\mu}  \label{velocity}
 \ee
  Explicitly
 \be
  \dot{t} = \frac{\mu  v^3}{M \sqrt{1-3 v^2+2 \mathfrak{q} v^3}} \qquad, \qquad  \dot{\phi}=\frac{\mu  \left(\mathfrak{q} v^3+1\right)}{\sqrt{1-3 v^2+2 \mathfrak{q} v^3}} \quad, \quad
  \dot{r}=\dot{\theta}=0
 \ee
The angular velocity is therefore
\be
\Omega={d\phi \over dt} ={\dot{\phi} \over \dot t} = {v^3 \over M (1+ \mathfrak{q} v^3) }
\ee

\subsection{Wave form and energy flux}

Putting together (\ref{tt}), (\ref{eqsource0}) and  (\ref{rt0}), one finds \cite{Mino:1997bx}
 \be
 Z_{\ell m} =
{{\pi   }\over { {\rm i} \omega B_{{\rm in}} }}
\left[  A_0 R_{{\rm in} }(r_0){+} A_1 R'_{{\rm in} }(r_0)
 {+}  A_2 R''_{{\rm in} }(r_0)    \right]_{\omega=m \Omega}
\label{zlma0}
\ee
with the $A_i$'s given by (\ref{ais}) in Appendix \ref{sourceterm}.
In terms of  $Z_{\ell m}$,  the wave form takes the form
\be
h_+  - {\rm i} h_{\times} ={2\pi \over r} \sum_{\ell,m} {Z_{\ell m} \over \omega^2} {}_{-2}S_{\ell m}(\chi)e^{ {\rm i} \omega (r^*-t)+{\rm i} m \phi }
\ee
and the gravitational luminosity (the amount of energy carried away per unit time) is given by
\be
{dE\over dt} =\mu^2 \sum_{\ell,m} { | Z_{\ell,m}|^2 \over 4\pi \omega^2}  \label{dedt}
\ee
 In the next section we will derive a PN expansion of these formulae for the Schwarzschild and Kerr geometries.

\section{CFT  gravity correspondence}

The basic idea behind the CFT gravity correspondence \cite{Bonelli:2021uvf,Bonelli:2022ten,Consoli:2022eey} relies on the fact, that
the type of differential equations governing the dynamics of black hole perturbations appear in the context of Liouville theory as conformal Ward identities \cite{Belavin:1984vu},  satisfied by certain five points conformal blocks in the limit of large central charge.
	
According to the AGT correspondence,  the degenerate conformal block is in turn related to the instanton partition function of a $N=2$ supersymmetric quiver gauge theory \cite{Fucito:2004gi,Alday:2009aq} with gauge group $SU(2)^2$. The large central charge limit corresponds to the Nekrasov-Shatashvili limit \cite{Nekrasov:2009rc} of the instanton partition function. In this section we collect the formulae \cite{Bonelli:2022ten,Consoli:2022eey} which will be relevent to our computation. We refer the reader to these references and Appendix \ref{cfttools} for details.

\subsection{Confluent Heun equation}

 The homogeneous part of the Teukolsky angular and radial equations (\ref{teukang}) can be both brought into the normal form
 \be
\left[ {d^2\over dz^2} +Q(z) \right] \Psi(z) = 0  \label{can}
\ee
with $z$ expressed in terms of the angular or radial variable and  \cite{Consoli:2022eey}
\be\label{Q21}
Q(z) =-\frac{{x}^2}{4
	z^4} + \frac{ {x}\, c }{z^3}+\frac{u  -\ft14 + x(\ft12-k_0) }
{(z-1) z^2}+\frac{ \ft14-k_0^2  }{(z-1)^2 z}+\frac{\ft14-p_0^2}{(z-1) z}
\ee
(\ref{Q21}) has two regular singularities at $z=\infty,1$ and an irregular one at $z=0$.  The solution can be written in terms of confluent Heun  functions with parameters  expressed in terms of $p_0$, $k_0$, $c$, $u$ and $x$.
 The parametrization (\ref{Q21}) is inspired by the correspondence with a five-point correlator in the Liouville theory that will be discussed below.

   \subsection*{ The angular equation}

 The angular equation can be written in the normal form (\ref{can}), by writing
 \be
 \chi={2\over z} - 1 \qquad, \qquad {}_{s} S_{\ell m} = {\Psi(z) \over \sqrt{1-\chi^2}}
 \ee
 and identifying\footnote{The superscript $\chi$ is to distinguish this angular equation from the radial one.}
  \be
x^\chi= 4 a \omega ~,~ c^\chi= s ~,~ p_0^\chi= \frac{s-m}{2}~,~ k^\chi_0=\frac{s+m}{2} ~,~ A= u^\chi-(s+\ft12)^2  +2 a  \omega (1-m)  - a^2 \omega ^2
\label{dicang}
  \ee

 \subsection*{ The radial equation}

The homogenous part of the radial equation can be written in the normal form (\ref{can}), by taking
 \be
 z={r_+-r_- \over r-r_-}   \qquad, \qquad  R_{\rm in} = (1-z)^{-{s+1\over 2}} \, z^{s}\,  \Psi(z)   \label{radialpsi}
 \ee
  and identifying
  \bea
x &=& 4 i \delta  \omega  \qquad , \qquad c= s{-}2 i M \omega \qquad , \qquad k_0 =
\ft{s}{2}- iP_+ \qquad , \qquad  p_0= \ft{s}{2} +iP_- \nn\\
     u & =& A{+}(s{+}\ft12)^2{+}2 \omega ( a m{-} i \delta ){+}  \omega ^2(a^2{-}12 M^2{-} 8 M \delta) \label{dicr}
     \eea
    with
    \bea
    r_{\pm}&=&M\pm \delta=M\pm \sqrt{M^2-a^2}\qquad , \qquad P_\pm=\frac{   a m{-}2  M \omega  r_\pm}{r_+-r_- }
    \eea

\subsection{The five-point conformal block}

 We consider the five point conformal blocks, $\Psi_\pm(z)$, with the insertion of a degenerate field at position $z$, two primaries at $\infty$, $1$, and  two colliding at $0$. $\Psi_\pm(z)$ are the independent solutions of (\ref{can}) \footnote{In the  paper $\pm$ and $\pm 1$ are interchangeable.} and are functions of the momenta, $p_0$, $k_0$, $c$, of the primary fields and are distinguished by the field exchanged in the intermediate channel. Using the AGT correspondence they can be written as
\bea
\Psi_\alpha(z)  =  \lim_{b \to 0}\, e^{{x\over 2z} }  \,  z^{{1\over 2} +\alpha \mathfrak{a}}     \,
\left(1{-} z  \right)^{ (2k_{0}{-}1 {-} b^2)( 2 k_{\rm deg} {+} 1 {+}  b^2 ) \over 2 b^2  }   {
	Z_{\rm inst}{}_{p_0}{}^{k_0}  {}_{\mathfrak{a}^{-\alpha}}{}^{ k_{\rm deg}}  {}_{\mathfrak{a}}  {}_c  (z,\ft{x}{z} ) \over
	Z_{\rm inst}{}_{p_0}{}^{k_0} {}_{\mathfrak{a}}   {}_c (x)   }  \label{psidef}
\eea
with  \cite{Flume:2002az,Bruzzo:2002xf,Fucito:2004gi,Alday:2009aq}
{\small
\bea
Z_{\rm inst}{}_{p_0}{}^{k_0} {}_{ \mathfrak{a} c} (q) &=&
	 \sum_{ W}  q^{ |W |}
	 \frac{ z^{\rm bifund}_{\emptyset,W} (p_0,\mathfrak{a},-k_0)    z^{\rm hyp}_{W} (\mathfrak{a},-c)
	 }{     z^{\rm bifund}_{W,W} (\mathfrak{a},\mathfrak{a},\ft{b^2+1}{2}) }   \label{zinstG2} \\
	Z_{\rm inst}{}_{p_0}{}^{k_0}  {}_{\mathfrak{a}^{-\alpha} } {}^{k_1} {}_{ \mathfrak{a} c} (q_1,q_2) &=&
	 \sum_{ Y,W}  q_1^{  |Y| } q_2^{ |W |}
	 \frac{ z^{\rm bifund}_{\emptyset,Y} (p_0,\mathfrak{a}^{-\alpha} ,-k_0) z^{\rm bifund}_{Y,W} (p_1,\mathfrak{a},-k_1)   z^{\rm hyp}_{W} (\mathfrak{a},-c)
	 }{  z^{\rm bifund}_{Y,Y} (\mathfrak{a}^{-\alpha},\mathfrak{a}^{-\alpha},\ft{b^2+1}{2})   z^{\rm bifund}_{W,W} (\mathfrak{a},\mathfrak{a},\ft{b^2+1}{2}) }
	\label{inst1}\eea
	}
 the instanton partition functions of $SU(2)$ and $SU(2)^2$ gauge theories with three flavours. The sums in (\ref{inst1}) run over the pairs of Young tableau $\{Y_\pm \}$, $\{W_\pm \}$ while $|Y|$, $|W|$ denote the total number boxes in each pair.
The functions  $z^{\rm bifund}_{Y,W }$, $z^{\rm hyp}_{Y_{2} }$ represent the contributions of a hypermultiplet transforming in the bifundamental and fundamental representation of the gauge group and are given by a product over the boxes of the Young tableau specified by $s=(i,j)$ (see Appendix \ref{cfttools} for details)
\bea
	 z^{\rm bifund}_{\Lambda, \Lambda' }( p , p' , m ) &=& \prod_{\beta,\beta'}
	    \prod_{s\in \Lambda_\beta}  \left[  E_{\Lambda_\beta,\Lambda'_{\beta'}}(\beta p{-}\beta' p',s)  {-} m  \right] \prod_{t\in \Lambda'_{\beta'}} \left[  { -}E_{\Lambda'_{\beta'},\Lambda_\beta}(\beta' p' {-}\beta p,t)  {-} m  \right]  \nn\\
z^{\rm hyp}_{\Lambda }(   \mathfrak{a} ,m ) &=& \prod_{\beta}
	    \prod_{s\in \Lambda_\beta}  \left[  -E_{\Lambda_\beta,\emptyset}(\beta \mathfrak{a},s)  +m  \right]
\label{inst2}\eea	
with
\bea
E_{\Lambda, \Lambda'}(x,s) &=& x- (\lambda_{ \Lambda' j}^T-i) + b^2 (\lambda_{ \Lambda i}-j+1)-\ft{b^2+1}{2}
\label{inst3}
\eea
where $\lambda_{ \Lambda i}$ is the number of boxes in the $i$-th row and $\lambda_{ \Lambda j}^T$ is the number of boxes in the $j$-th column of the tableau $\Lambda$.
The conformal field theory variables entering in the five-point conformal block are related to the $SU(2)^2$  gauge theory couplings, $\Omega$-background parameters  and masses via the dictionary
	\bea
	q_1& =& z \quad , \quad q_2={x\over z} \quad, \quad  \epsilon_1=1 \quad, \quad \epsilon_2=b^2 \nn\\
	m_1 &=& k_0+p_0 \quad , \quad m_2=k_0-p_0 \quad ,\quad m_3=-c  \quad , \quad m_{\rm bifund} =-k_{\rm deg}
	\eea
Similar relations hold between the four-point conformal block and the $SU(2)$ gauge theory with coupling now given by  $q=q_1 q_2=x$.
 On the other hand
\bea
 k_{\rm deg}=\ft{1}{2} + b^2    \quad, \quad  p_s^\alpha = p_s+\frac{\alpha b^2}{2}
\label{vnull}
\eea
are the momenta of the degenerate insertion and of its OPE  with a primary field with momentum $p_s$\cite{Belavin:1984vu}. We notice that each factor in (\ref{psidef}) is divergent in the limit $b \to 0$ but the product is finite. The result is given by a double series expansion
\bea
\Psi_\alpha(z) &=&  z^{{1\over 2} +\alpha \mathfrak{a}} \left[  1+z \frac{ k_0^2- p_0^2+
	\mathfrak{a}^2-\ft14}{1+2 \alpha  \mathfrak{a}} -\frac{c {x} }{z \left(1-2 \alpha  \mathfrak{a}\right)} \right . \nn\\
&&   \left.  -
c {x}  \left(\frac{4 \left(k_0^2-p_0^2 \right) \left(2 \alpha  \mathfrak{a}+3\right)-(3-2\alpha \mathfrak{a})(1+2\alpha \mathfrak{a})^2  }{4 \left(1-2 \alpha  \mathfrak{a}\right) \left(2
	\alpha  \mathfrak{a}+1\right){}^2}\right)  \right] +\ldots \label{psialpha}
\eea
Finally the parameter $\mathfrak{a}$ labels the momenta of the Liouville field exchanged in the intermediate channel and characterizes the monodromy of the gravity solutions around $z=0$
\be
\Psi_\alpha\left( e^{2\pi {\rm i} } z \right) =e^{2\pi {\rm i} \left({1\over 2}+\alpha \mathfrak{a}\right) }\Psi_\alpha\left(  z \right)
\ee
In the gauge theory framework $\mathfrak{a}$ represents the scalar vacuum expectation value (the fundamental Seiberg-Witten period) and can be determined by solving
\be
u=\mathfrak{a}^2 -  {x}  \partial_{x} {\cal F}_{\rm inst} (\mathfrak{a})    \label{uf}
\ee
as a series in $x$ with
\bea
{\cal F}_{\rm inst} &=&  \lim_{b\to 0} b^2   \ln \left[  Z_{\rm inst}{}_{p_0}{}^{k_0} {}_{\mathfrak{a}}   {}_c (x) \right]
=  \frac{x}{2} \left[ (1+c-2k_0)-{4c \left( k_0^2- p_0^2\right) \over 1-4 \mathfrak{a}^2} \right]
+  \ldots   \label{u4}
\eea
The result reads
\be
\mathfrak{a}=\sqrt{u}+ {x \over \sqrt{u}} \left[ \ft{1}{4} \left(1+c-2 k_0\right) + \frac{c \left(  k_0^2- p_0^2\right)}{4 u-1} \right]
+  \ldots
\ee
\subsection{Braiding and fusion connection formulae }

According to the gauge gravity correspondence, the double instanton series (\ref{psialpha}) describes the PN expansion of the gravitational waveform in the limit where both $z$ and $x/z$ are small. In the gravity variables, see (\ref{radialpsi}-\ref{dicr}), this corresponds to the region where the distance from horizons are large but still much  smaller than the wavelength of the perturbation
\be
\omega (r - r_-) \ll 1 \qquad  {\rm and} \qquad   r \gg r_+ \label{nearzone}
\ee
We will refer to this region as the {\it near zone}. We notice that what it is typically called the near zone is the region where $\omega (r-r_+) \ll 1$.
The  PN expansion requires in addition that $r \gg r_+$, so we could better call the region described by (\ref{nearzone}), the asymptotically far   near zone, but for simplicity we will refer to it  as the near zone.
 Moreover, we will further divide this zone into two patches depending on whether $x/z \ll z$ (int) or $z \ll x/z$ (ext). In these two extreme cases the instanton series (\ref{psialpha}) can be resummed  and written in terms of the hypergeometric  functions
	\bea
\widehat{\widehat{\psi}}_\alpha (z) & \equiv& \lim_{\ft{x}{z}\to 0}	\Psi_\alpha(z)   = z^{{1\over 2} {+}\alpha \mathfrak{a}}    (1{-}z)^{ {1\over 2} {+}k_0}  {}_2 F_1( \ft12{+}k_0{+}p_0{+}\alpha \mathfrak{a}, \ft12{+}k_0{-}p_0+\alpha \mathfrak{a},1{+}2 \alpha \mathfrak{a} ,z)  \nn\label{fx00} \\
 \widecheck{\psi}_\alpha (\ft{x}{z}) &  \equiv& x^{-{1\over 2}-\alpha \mathfrak{a} }  \lim_{z \to 0 } \Psi_\alpha(z)   =  \left(\ft{z}{x}\right)^{{1\over 2} {+}\alpha \mathfrak{a}} e^{-{x\over 2 z}} {}_1 F_1( \ft12-c-\alpha \mathfrak{a},  1-2 \alpha \mathfrak{a} ,\ft{x}{z})
	\label{fx0}
	\eea
In these patches the effective potential can be approximated as	
\bea
Q_{\rm int} (z)  & \approx & \frac{\mathfrak{a}^2-\ft14}{(z-1) z^2}+\frac{\frac{1}{4}-k_0^2 }{(z-1)^2 z}+\frac{\frac{1}{4}-p_0^2}{(z-1) z}  \nn\\
Q_{\rm ext}(z) &\approx &-\frac{{x}^2}{4z^4} + \frac{ {x}\, c }{z^3}+\frac{\mathfrak{a}^2-\ft14}{(z-1) z^2}
\eea	
Using the standard hypergeometric connection formulae these functions can be alternatively written as
\bea	
\widehat{\widehat{\psi}}_\alpha (z) &=& F_{\alpha \alpha''} \widehat{\psi}_{\alpha''}(z) \nn\\
\widecheck{\psi}_\alpha (\ft{x}{z}) & = & B^{\rm conf}_{\alpha \alpha'} \widetilde{\psi}_{\alpha'}( \ft{x}{z}) \label{hypbf}
\eea	
with
\bea
\widehat{\psi}_{\alpha''} (z)&  \equiv & z^{ \frac{1}{2} + \mathfrak{a}} \,   (1{-}z)^{\frac{1}{2}{+}\alpha'' k_0} {}_2F_1\left(\ft{1}{2}{+} \alpha'' k_0{+}p_0{+} \mathfrak{a},
\frac{1}{2}{+} \alpha'' k_0{-}p_0{+} \mathfrak{a};1 {+}2\alpha'' k_0 ;1{-} z\right)  \nn\\
\widetilde{\psi}_{\alpha'} (\ft{z}{x})  &  \equiv &   \left(\ft{z}{x}\right)^{ 1+ \alpha' c}   e^{{ \alpha' x\over 2z}}    \, _2F_0\left(\ft{1}{2}{+}\alpha' c{+} \mathfrak{a};\ft{1}{2}{+}\alpha' c{-} \mathfrak{a} ;\ft{\alpha' z}{ x}\right)  \label{hypf}
\eea
and
\bea
F_{\alpha \alpha''  }  &=& \frac{\Gamma (1+2  \alpha \mathfrak{a} )
	\Gamma \left(-2  \alpha''  k_0 \right)}{\Gamma \left(\frac{1}{2}+p_0+\alpha \mathfrak{a}-\alpha'' k_0
	\right) \Gamma \left(\frac{1}{2}-p_0+\alpha \mathfrak{a}-\alpha'' k_0 \right)} \nn\\
F^{-1}_{\alpha'' \alpha }  &=& \frac{\Gamma (1+2  \alpha'' k_0)
	\Gamma \left(-2  \alpha \mathfrak{a}\right)}{\Gamma \left(\frac{1}{2}+\alpha'' k_0+p_0-\alpha \mathfrak{a}
	\right) \Gamma \left(\frac{1}{2}+\alpha'' k_0-p_0-\alpha \mathfrak{a}\right)} \nn\\
B^{\rm conf}_{\alpha \alpha'}  &=&  (\alpha')^{-{1\over 2}+\alpha p_2+\alpha' c}   \frac{   \Gamma \left(1-2  \alpha \mathfrak{a}\right) }
{\Gamma \left(\frac{1}{2} -\alpha \mathfrak{a}-\alpha' c \right) }  \label{braidingBconf}
\eea
$\widehat{\psi}_{\alpha}$, $\widetilde{\psi}_{\alpha}$ arise as limits of the conformal blocks
$  \widehat\Psi_{\alpha} $ and   $  \widetilde\Psi_{\alpha} $ describing the {\it near horizon} and {\it far zone} respectively.
The latters correspond to different OPE expansion of the same five-point correlator and are related to each other by the CFT braiding and fusion relations
\cite{Bonelli:2021uvf,Bonelli:2022ten,Consoli:2022eey}
\be
\widehat{\Psi}_{\alpha''} (z)=  F^{-1}_{\alpha'' \alpha}   \Psi_\alpha (z) \qquad  , \qquad    \Psi_{\alpha}(z) = B^{\rm conf}_{\alpha \alpha'}  \tilde{\Psi}_{\alpha'} (z)
\ee
where the fusion and braiding matrices are given in (\ref{braidingBconf}).
Any solution of the homogenous differential equation can be written in the alternative ways
\be
\Psi  (z) =\sum_{\alpha=\pm}  c_\alpha \widehat{\Psi}_\alpha(z)=
\sum_{\alpha=\pm}  \hat{c}_\alpha \widehat{\Psi}_\alpha(z)=\sum_{\alpha=\pm}  \tilde{c}_\alpha \widetilde{\Psi}_\alpha(z)
 \label{pinc2}
\ee
with
\bea
\widehat{c}_{\alpha'} =F_{\alpha' \alpha}\,  c_\alpha  \qquad ,\qquad  \tilde{c}_{\alpha'} = c_{\alpha} \,  x^{{1\over 2}+\alpha \mathfrak{a}}  B^{\rm conf}_{\alpha \alpha'}
\eea
 For example, the solution $\Psi_{\rm in} (z)$ satisfying the incoming boundary conditions
 \bea
\Psi_{\rm in} (z)   & \approx &
\left\{
\begin{array}{ccc}
 (1-z)^{{1\over 2} -k_0}  &   & \qquad\quad z\to 1   \\
	C_-    e^{ -\frac{{x}}{2z}}  z^{1-c}  + C_+  e^{ \frac{{x}}{2z}}   z^{1+c}  &~~~~~~~~~~~~~~~~~~~~   & \qquad\quad z\to  0   \\
\end{array}
\right.
 \label{bcpsi2}
\eea
 can be written in the three regions as
\be
 \Psi_{\rm in}  =\left\{
\begin{array}{lllll}
  \widehat{\Psi}_{-} (z) & ~~~ & {\rm Near~ horizon}  &~~&  z \approx 1 \\
	 F^{-1}_{-, \alpha}  \Psi_\alpha(z) & &  {\rm Near~ zone}  && 1\gg z \gg x \\
	 x^{{1\over 2}+\alpha \mathfrak{a}}  F^{-1}_{- ,\alpha}  B^{\rm conf}_{\alpha\alpha'}   \widetilde{\Psi}_{\alpha'} (z) & &  {\rm Far ~away} & & 1 \gg x  \gg z
\end{array}\right.
\label{braidfus}
\ee
For the PN expansion we will use
\be
\Psi_{\rm in}=\sum_\alpha   F^{-1}_{-,\alpha} \Psi_\alpha
\ee
On the other hand, the asymptotics at  infinity is obtained by expanding the conformal blocks   $ \widetilde\Psi_{\alpha'}(z) $ near $z=0$.
In this limit, the five-point conformal block factorizes into a four-point function depending only on $x$,
 times a function of $z$ (see Appendix \ref{cfttools} for details).  The latter is obtained by expanding the hypergeometric function in the second line of (\ref{hypf}) in the limit where $z\ll x$.   One finds
 \bea
\Psi_{\rm in} (z)
\underset{z\to 0}{\approx} &&
\sum_{\alpha,\alpha'}
				F^{-1}_{-, \alpha}
				B^{\rm conf}_{\alpha  \alpha'}   \, e^{{\alpha' {x} \over 2 z } } z^{ 1+\alpha' c} x^{ -{1\over 2} +\alpha' c+\alpha \mathfrak{a}}  \lim_{b\to 0}
				  { Z_{\rm inst}{}_{p_0}{}^{k_0} {}_{ \mathfrak{a}^{-\alpha}  c^{-\alpha'} } (-x) \over Z_{\rm inst}{}_{p_0}{}^{k_0} {}_{ \mathfrak{a} c} (-x) } =\nonumber \\
				  &=&\sum_{\alpha,\alpha'}
				  F^{-1}_{-, \alpha}
				  B^{\rm conf}_{\alpha  \alpha'} \,e^{{\alpha' {x} \over 2 z } }z^{ 1 +\alpha' c} \,   x^{ -{1\over 2} -\alpha' c+\alpha \mathfrak{a}}e^{-\frac{1}{2}  (\alpha' \partial_c +\alpha \partial_{\mathfrak{a}}){\cal F}_{\rm inst}(-x)}
				 \label{connection}
\eea
Comparing against (\ref{bcpsi2}) one finds
\be
C_{\alpha'} ({x} ) = \sum_{\alpha}    F^{-1}_{-, \alpha}
				  B^{\rm conf}_{\alpha  \alpha'} \,e^{{\alpha' {x} \over 2 z } } \,   x^{ -{1\over 2} -\alpha' c+\alpha \mathfrak{a}}e^{-\frac{1}{2}  (\alpha' \partial_c +\alpha \partial_{\mathfrak{a}}){\cal F}_{\rm inst}(-x)}
				    \label{calphap}
\ee

\section{Schwarzschild  BH  }

In this section, we  deal with the case of the Schwarzschild geometry. We consider a particle of mass $\mu$ moving in a circular orbit of  radius $r_0$  along the equatorial  plane.
From (\ref{velocity}) and (\ref{ptpp}) one finds
\be
\dot{x}^\mu=\left( -{P_t  \, r_0^2 \over \Delta (r_0)  }   , 0 , 0,    {P_\phi \over r_0^2 }   \right)= {\mu   \over   \sqrt{1-3v^2}}  \left( 1  ,0,0, \Omega  \right) \label{geosch}
\ee
with
\be
v^2=\frac{M}{r_0} \qquad,\qquad \Omega= {v^3\over M}
\ee
 For circular orbits
\be
  \omega  =m\Omega={m v^3 \over M} \label{geosch2}
\ee

\subsection{CFT gravity dictionary}

 In the Schwarschild case the angular equation can be solved in terms of the spin weighted
 spheroidal harmonics    $ {}_{-2} S_{\ell m}(\cos\theta)$ with eigenvalues
 \be
 A=(\ell+2)(\ell-1)
 \ee
  The radial  equation reduces to
\be
\Delta_r (r)^2 {d\over dr} \left[  {R_{\ell m}'(r) \over \Delta_r(r) } \right]  +\left(\frac{ \omega^2 r^4 +4{\rm i}  (r-M) \omega r^2 }{\Delta_r(r)}-8{\rm i}  \omega r- (\ell+2)(\ell-1)  \right)R_{\ell m}(r)  =T_{\ell m}(r)\label{teuk1}
\ee
with
\bea
\Delta_r(r) &=&  r(r -2 M)
\eea
As we said earlier the radial equation can be put into the normal form (\ref{can}) with
\be
z = {2M\over r} \qquad \qquad    R(r) ={(1-z)^{1\over 2} \over z^2  }   \Psi(z) \label{rzdic}
\ee
and
\be
Q=\frac{4 \ell(\ell+1) z^2(z-1) +16 M^2 \omega ^2-3 z^4+\ii M z \omega(48z-32)}{4 (z-1)^2 z^4} \label{qr}
\ee
Comparing against  (\ref{Q21}) one finds the gauge gravity dictionary
\bea
x  &=& 4 i   \omega M  \qquad , \qquad  z={2M\over r} \label{dicrsch}  \\
c &=&  {-}2{-}2 i M \omega \quad ,\quad  p_0 =  -1   \quad , \quad k_0 =
{-}1{+} 2 i M \omega  \quad ,\quad
u =  (\ell{+}\ft12)^2 {-}2 i M \omega {-}20 \omega^2 M^2  \nn
\eea

\subsection{The PN expansion of $R_{\rm in}$ and $B_{\rm in}$ }

The incoming solution $R_{\rm in}(r)$ and the asymptotic coefficient $B_{\rm in}$ are given by the CFT formulae
 \bea
 R_{\rm in} &=&  z^{-2} (1-z)^{1\over 2}  \sum_{\alpha = \pm} F^{-1}_{- \alpha }  \Psi_\alpha(z)\nn\\
B_{\rm in} &=& 2M  C_- = 2M \sum_{\alpha}     F^{-1}_{- \alpha}
B^{\rm conf}_{\alpha  -}   x^{{-}{1\over 2}{+}c{+}\alpha \mathfrak{a}}  e^{{ 1\over 2}  ( \partial_c-\alpha \partial_{\mathfrak{a}}) {\cal F}_{\rm inst}    }   \label{binexp}
\eea
We notice that the PN expansion is largely dominated by the $\Psi_-$ contribution, since $\Psi_+$ is suppressed by an extra $z^{2\mathfrak{a}}\sim r^{-2\ell-1}$ factor. The two solutions start to mix at order $v^{10}$ in the PN expansion for $\ell=2$ modes.
The mixing is codified by the  tidal function ${\cal L}=F_{-+}^{-1}/F_{--}^{-1}$ computing the ratio response/source. We will study this function later in this paper.  Here we just notice that the leading contribution  for a Schwarschild BH's vanishes in the static limit $\omega\to 0$\footnote{Here the limit is computed by sending first $\omega \to 0$ and then $\ell \to \mathbb{Z}$.  The same result is obtained
 by taking the limits in the opposite order: using first (\ref{p2final}) with  $\ell\in\mathbb{Z}$ and sending then $\omega\to 0$.
  In this case,  both $\Psi_{\pm}$ contribute to  order  $z^{2 \widehat{\ell}+1}$  but they cancel against each other.
 A similar cancellation occurs in the case of the Kerr geometry, where both $\Psi_{\pm}$  exhibit a pole in $\omega$ at order
 $z^{2\widehat{\ell}+1}$ but they cancel against each other \cite{Charalambous:2021mea}. We thank C. Iossa for pointing this reference to us. }.

 We introduce the  short hand notation
\be
\epsilon=2\omega M \sim v^3     \qquad , \qquad \kappa=\omega r \sim v
\label{epsilonkappa} \ee
In term of these variables (\ref{dicrsch}) becomes
\bea
x  &=& 2 i  \epsilon  \qquad , \qquad  z=\frac{\epsilon}{\kappa}\quad \label{dicr1}  \\
c &=&  {-}2{-} i \epsilon \quad ,\quad  p_0 =  -1   \quad , \quad k_0 =
{-}1{+}  i \epsilon  \quad ,\quad
u =  (\ell{+}\ft12)^2 {-} i \epsilon{-}5\epsilon^2  \nn
\eea
and
\bea
R_{\rm in} &=&  F^{-1}_{--} \left( {\epsilon \over \kappa} \right)^{-{3\over 2}-\mathfrak{a}} \sum_{n=0}^{\infty}  R_{n} + (\mathfrak{a}\to {-}\mathfrak{a}) \nn\\
B_{\rm in} &=& 2M  F^{-1}_{--}  (-2 {\rm i} \epsilon )^{-{5\over 2}-{\rm i} \epsilon-\mathfrak{a}}  {\Gamma(1+2 \mathfrak{a})\over \Gamma(-\ft32-{\rm i} \epsilon+\mathfrak{a}) } \sum_{n=0}^{\infty}  B_{3n}+ (\mathfrak{a}\to {-}\mathfrak{a}) \label{rbin}
\eea
 where $B_n$ and ${R}_n$ scale as $v^n$ in the PN limit. $\mathfrak{a}$ is  obtained  by solving (\ref{uf}) order by order in $\omega$ and using the dictionary (\ref{dicr1}) leading to
 \be
\mathfrak{a}=\ell +\ft{1}{2}-\epsilon^2\frac{\left(15 \ell^4+30 \ell^3+28 \ell^2+13 \ell+24\right) }{2 \ell (\ell+1) (2 \ell-1) (2 \ell+1) (2 \ell+3)}+ O(\epsilon^4) \label{p2final}
\ee
 This formula reproduces (5.2) of \cite{Mano:1996vt} for $s=-2$ after the identification $\nu=\mathfrak{a}-\ft12$\footnote{The formula for arbitrary s is reproduced  in a similar way using the dictionary (\ref{dicrsch}).}.
 The first  non-trivial coefficients $B_n$'s  are
\bea
 && B_0= 1 \nn\\
 && B_3 =\frac{ {\rm i}  \epsilon }{2} \nn\\
  && B_6=- \epsilon ^2\frac{\left(32 \mathfrak{a}^6-296 \mathfrak{a}^4-900 \mathfrak{a}^3+136 \mathfrak{a}^2+225
   \mathfrak{a}+128\right) }{64 \left(\mathfrak{a}^2-1\right){}^2 \left(4
   \mathfrak{a}^2-1\right)} \nn\\
  && B_9= -{\rm i}\epsilon ^3  \frac{ \left(8 \mathfrak{a}^4+32 \mathfrak{a}^2-675 \mathfrak{a}-40\right)  }{384
   \left(\mathfrak{a}^2-1\right){}^2}
\eea
while the $R_n$'s are given by
{\small
\bea
 && R_0= 1 \nn\\
&&R_1= \frac{4 i  \kappa }{1{+}2 \mathfrak{a}}\nn\\
&&R_2=\frac{\epsilon  \left({-}3{-}2 \mathfrak{a}\right)}{4
   \kappa }{-}\frac{\kappa ^2 \left(17{+}2 \mathfrak{a}\right)}{4{+}12 \mathfrak{a}{+}8 \mathfrak{a}^2}\nn\\
 &&R_3=
   {-}\frac{i \epsilon  \left(5{+}2 \mathfrak{a}\right)}{1{+}2 \mathfrak{a}}{-}\frac{i \kappa ^3 \left(7{+}2
   \mathfrak{a}\right)}{3{+}11 \mathfrak{a}{+}12 \mathfrak{a}^2{+}4 \mathfrak{a}^3} \label{rfinaln} \\
  &&R_4=\frac{\kappa ^4 \left(163{+}72 \mathfrak{a}{+}4 \mathfrak{a}^2\right)}{32 \left(1{+}\mathfrak{a}\right)
   \left(2{+}\mathfrak{a}\right) \left(1{+}2 \mathfrak{a}\right) \left(3{+}2 \mathfrak{a}\right)}{+}\frac{\epsilon  \kappa
   \left(87{+}16 \mathfrak{a}{+}4 \mathfrak{a}^2\right)}{16{+}48 \mathfrak{a}{+}32 \mathfrak{a}^2}{+}\frac{\epsilon ^2 \left(9{+}18
   \mathfrak{a}{-}4 \mathfrak{a}^2{-}8 \mathfrak{a}^3\right)}{\kappa ^2 \left(64{-}64 \mathfrak{a}\right)}\nn\\
 &&R_5=  \frac{i \kappa ^5 \left(11{+}2
   \mathfrak{a}\right)}{8 \left(1{+}\mathfrak{a}\right) \left(2{+}\mathfrak{a}\right) \left(1{+}2 \mathfrak{a}\right) \left(3{+}2
   \mathfrak{a}\right)}{+}\frac{i \epsilon  \kappa ^2 \left(15{+}4 \mathfrak{a}^2\right)}{4 \left(3{+}11 \mathfrak{a}{+}12
   \mathfrak{a}^2{+}4 \mathfrak{a}^3\right)}{+}\frac{i \epsilon ^2 \left(61{-}72 \mathfrak{a}{+}8 \mathfrak{a}^2{+}32 \mathfrak{a}^3{+}16
   \mathfrak{a}^4\right)}{16 \kappa  \left({-}1{+}\mathfrak{a}\right) \left({-}1{+}2 \mathfrak{a}\right) \left(1{+}2
   \mathfrak{a}\right)} \nn\\
   &&R_6=
   {-}\frac{\kappa ^6 \left(339{+}104 \mathfrak{a}{+}4 \mathfrak{a}^2\right)}{384
   \left(1{+}\mathfrak{a}\right) \left(2{+}\mathfrak{a}\right) \left(3{+}\mathfrak{a}\right) \left(1{+}2 \mathfrak{a}\right)
   \left(3{+}2 \mathfrak{a}\right)}{+}\frac{\epsilon  \kappa ^3 \left(1405{+}806 \mathfrak{a}{-}180 \mathfrak{a}^2{-}24
   \mathfrak{a}^3\right)}{384 \left(1{+}\mathfrak{a}\right) \left(2{+}\mathfrak{a}\right) \left(1{+}2 \mathfrak{a}\right)
   \left(3{+}2 \mathfrak{a}\right)}  \nn\\
   &&{+}\frac{\epsilon ^3 \left(15{+}4 \mathfrak{a}{-}64 \mathfrak{a}^2{-}16 \mathfrak{a}^3{+}16
   \mathfrak{a}^4\right)}{\kappa ^3 \left(768{-}768 \mathfrak{a}\right)}{-}\frac{\epsilon ^2 \left({-}4239{+}3641
   \mathfrak{a}{+}456 \mathfrak{a}^2{-}8452 \mathfrak{a}^3{+}4624 \mathfrak{a}^4{+}112 \mathfrak{a}^5{-}256 \mathfrak{a}^6{+}64 \mathfrak{a}^7\right)}{256
   \left({-}1{+}\mathfrak{a}\right){}^2 \left(1{+}2 \mathfrak{a}\right){}^2 \left({-}1{+}\mathfrak{a}{+}2
   \mathfrak{a}^2\right)} \nn
      \eea
The results up to $n=10$ are displayed in Appendix \ref{mathrin}.
Computing the ratio $R_{\rm in}/ B_{\rm in}$, and using (\ref{p2final}) to express $\mathfrak{a}$ in terms of $\ell$,  one finds
\be
\mathfrak{R}_{\rm in}={  R_{\rm in} \over 2 {\rm i} \omega B_{\rm in} } =(-2{\rm i})^{ \widehat{\ell}+2+{\rm i} \epsilon } \epsilon^{{\rm i} \epsilon} \kappa^{\widehat{\ell}+2}
{ \Gamma(\widehat{\ell}-1-{\rm i} \epsilon ) \over   \Gamma(2\widehat{\ell}+2 ) } \sum_{n=0}^\infty \mathfrak{R}_n
\label{gothicR0}\ee
with $\hat\ell=\mathfrak{a}-\frac12$ and
{\small
\bea
&&\mathfrak{R}_0= 1 \nn\\
&&\mathfrak{R}_1=\frac{2 i  \kappa }{1+\ell}\nn\\
&&\mathfrak{R}_2= \frac{\left(-1-\frac{\ell}{2}\right) \epsilon }{\kappa }-\frac{(9+\ell) \kappa ^2}{6+10 \ell+4 \ell^2}\nn\\
&&\mathfrak{R}_3= -\frac{i (7+3 \ell)
	\epsilon }{2 (1+\ell)}-\frac{i (4+\ell) \kappa ^3}{6+13 \ell+9 \ell^2+2 \ell^3}\nn\\
&&\mathfrak{R}_4= \frac{\left(2+\ell-2 \ell^2-\ell^3\right) \epsilon ^2}{(4-8 \ell) \kappa ^2}+\frac{\left(36+13
	\ell+\ell^2\right) \epsilon  \kappa }{12+20 \ell+8 \ell^2}+\frac{\left(50+19 \ell+\ell^2\right) \kappa ^4}{8 (1+\ell) (2+\ell) (3+2 \ell) (5+2 \ell)}\nn\\
&&\mathfrak{R}_5=\frac{i \left(4-6 \ell+11
	\ell^2+13 \ell^3+4 \ell^4\right) \epsilon ^2}{4 \ell (1+\ell) (-1+2 \ell) \kappa }+\frac{i \left(26+13 \ell+3 \ell^2\right) \epsilon  \kappa ^2}{4 (2+\ell) \left(3+5 \ell+2 \ell^2\right)}+\frac{i
	(6+\ell) \kappa ^5}{4 (1+\ell) (2+\ell) (3+2 \ell) (5+2 \ell)}\nn\\
&&\mathfrak{R}_6=-\frac{\left(\mathfrak{a}360+624 \ell+859 \ell^2+1016 \ell^3+662 \ell^4+234 \ell^5+43 \ell^6+2 \ell^7\right) \epsilon ^2}{8 \ell (1+\ell)^2
	(-1+2 \ell) (3+2 \ell)^2}+\frac{\ell \left(-4-4 \ell+\ell^2+\ell^3\right) \epsilon ^3}{(24-48 \ell) \kappa ^3}\nn\\
&&-\frac{\left(260+236 \ell+75 \ell^2+3 \ell^3\right) \epsilon  \kappa ^3}{48 (1+\ell)
	(2+\ell) (3+2 \ell) (5+2 \ell)}-\frac{\left(98+27 \ell+\ell^2\right) \kappa ^6}{48 (1+\ell) (2+\ell) (3+2 \ell) (5+2 \ell) (7+2 \ell)}
\label{gothicR}\eea	
  The remaining contributions up to $v^{10}$ are displayed in Appendix \ref{mathrin}.

\subsection{Wave form and gravitational luminosity}

 Plugging the  stress energy tensor associated to the circular orbits (\ref{geosch}), (\ref{geosch2}) into (\ref{zlma0}), and specifying to the Schwarzschild geometry ($a=0$, $\delta=M$)
 one finds (\ref{zlma0}) where \cite{Poisson:1993vp}
 \bea
 A_0 &=& {1\over 2 r_0^2}  \left(i \omega r_0 f_0^{-2}   b_{2\ell m} \left(\frac{M}{r_0}{-}\frac{1}{2} i\omega  r_0  { -} 1\right) {+}2 i \left(1{+}\frac{i  \omega r_0 }{2 f_0}\right)
 b_{1\ell m}{+}b_{0\ell m}\right)\nn\\
 A_1 &=& -{1\over 2 r_0}   \left(i b_{1 \ell m}-\left(1+\frac{i \omega r_0  }{f_0}\right) b_{2\ell m}\right) \nn\\
 A_2&=& -\ft14  b_{2\ell m}
 \label{asch}\eea
  with $f_0=1-2M/r_0$ and
   \bea
   b_{0\ell m} &=& -\ft{ P_t}{2 f_0} \sqrt{(\ell-1)\ell(\ell+1)(\ell+2)} {}_{0}Y_{\ell m}(\ft{\pi}{2},0)   \nn\\
   b_{1\ell m} &=& \ft{P_\phi}{r_0} \sqrt{(\ell-1)(\ell+2)} {}_{-1}Y_{\ell m}(\ft{\pi}{2},0) \nn\\
   b_{2\ell m} &=&  P_\phi\Omega   {}_{-2}Y_{\ell m}(\ft{\pi}{2},0)
  \label{bsch} \eea
The spin-weighted spherical harmonics are
\be
{}_s Y_{\ell m}(\theta,\phi)={e^{i m \phi } \over 2 \sqrt{\pi} }  \sin ^{2 l}\left(\frac{\theta }{2}\right)  \sqrt{\frac{(2 l+1) (l-m)! (l+m)!}{(l-s)! (l+s)!}} 
\sum_{r=0}^{\ell-s}  (-)^{m-r+l-s} (^{l-s}_r) (^{\ell+s}_{r+s-m}) \cot (\ft{\theta}{2})^{2r+s-m} \label{ylm} 
\ee

Plugging (\ref{gothicR0}), (\ref{asch}), (\ref{bsch}) into (\ref{zlma0}) one finds
 \be
 Z_{\ell m}=-16\sqrt{\frac{\pi}{5}}  {v^8\over M^2} z_{\ell m}
 \ee
 with (up to order $v^4$)
 \bea
 z_{2,2} &=& 1-\frac{107 v^2}{42}+\frac{1}{3} i v^3 \left[-17+12 \gamma -6 i \pi +36 \log (2 v ) \right]-\frac{2173 v^4}{1512} +\ldots \nn\\
 z_{2,1} &=& -\frac{i v}{12}+\frac{17 i v^3}{336} +\frac{1}{36} v^4 (-10+6 \gamma -3 i \pi +12 \log (2)+18 \log (v))+\ldots  \nn\\
 z_{3,3} &=& \frac{27}{16} i \sqrt{\frac{15}{14}} v-\frac{27}{4} i \sqrt{\frac{15}{14}} v^3 -\frac{27}{32}
   \sqrt{\frac{3}{70}} v^4 (-127+60 \gamma -30 i \pi +60 \log (12 v^3) )+\ldots  \nn\\
   z_{3,2} &=& \frac{1}{3} \sqrt{\frac{5}{7}} v^2-\frac{193 v^4}{54 \sqrt{35}}-\frac{1451 v^6}{2376 \sqrt{35}}+\frac{2}{9} \sqrt{\frac{5}{7}} v^5 (-13 i+6 i   \gamma +3 \pi +18 i \log (2v) )+\ldots  \nn\\
   z_{3,1} &=& -\frac{i v}{48 \sqrt{14}}+\frac{i v^3}{18 \sqrt{14}}+\frac{v^4 (-127+60 \gamma -30 i \pi +60 \log (4 v^3) )}{1440 \sqrt{14}}+\ldots  \nn\\
   z_{4,4} &=&-\frac{32}{9} \sqrt{\frac{5}{7}} v^2+\frac{9488 v^4}{99 \sqrt{35}}+\ldots   ~ , ~
   z_{4,3} = \frac{81 i v^3}{16 \sqrt{70}}+\ldots  ~,~
   z_{4,2}=\frac{\sqrt{5} v^2}{63}-\frac{437 v^4}{1386 \sqrt{5}}+\ldots  \nn\\
    z_{4,1}&=& -\frac{i v^3}{336 \sqrt{10}}+\ldots  ~,~
    z_{5,5}=-\frac{15625 i v^3}{384 \sqrt{66}}+\ldots  ~,~
    z_{5,4} = -\frac{128 v^4}{9 \sqrt{165}}+\ldots  \nn\\
    z_{5,3} &=&\frac{81}{128} i \sqrt{\frac{3}{110}} v^3+\ldots ~,~
    z_{5,2} = \frac{2 v^4}{27 \sqrt{55}}+\ldots  ~,~
    z_{5,1} =-\frac{i v^3}{1152 \sqrt{385}}+\ldots
 \eea
and $\gamma$ is the Euler gamma. From (\ref{dedt}) one finds for the gravitational luminosity
{\small
\bea
&& {dE\over dt} = {32  \mu^2 v^{10} \over 5 M^2} \left[ 1-\frac{1247 v^2}{336}+4 \pi  v^3-\frac{44711 v^4}{9072}-\frac{8191 \pi  v^5}{672} \right. \nn\\
&&  \left.+v^6 \left(\frac{6643739519}{69854400}-\frac{1712 \gamma
}{105}+\frac{16 \pi ^2}{3}-\frac{3424 \log (2)}{105}-\frac{1712 \log (v)}{105}\right)+ \frac{16285 \pi  v^7}{504}\right.\nn\\
&& +v^8 \left(-\frac{323105549467}{3178375200}+\frac{232597 \gamma
   }{4410}-\frac{1369 \pi ^2}{126}+\frac{39931 \log (2)}{294} -\frac{47385 \log
   (3)}{1568}+\frac{232597 \log (v)}{4410}\right) \nn\\
   &&+\pi  v^9
   \left(-\frac{265978667519}{745113600}+\frac{6848 \gamma }{105}+\frac{13696 \log
   (2)}{105}+\frac{6848}{105} \pi  \log (v)\right)+v^{10}
   \left({-}\frac{2500861660823683}{2831932303200} \right. \nn\\
   && \left.\left. {+}\frac{916628467 \gamma
   }{7858620}  {-}\frac{424223 \pi ^2}{6804}{-}\frac{83217611 \log (2)}{1122660}{+}\frac{47385
   \log (3)}{196}{+}\frac{916628467 \log (v)}{7858620}\right) +\ldots \right]
\eea
}
in agreement with  previous results in the literature \cite{Tagoshi:1993dm,Tagoshi:1994sm,Shibata:1994jx,Tanaka:1996lfd,Mino:1997bx,Fujita:2010xj,Fujita:2011zk,Fujita:2012cm}.

\section{Kerr  BH  }

In the case of the Kerr geometry both the radial and the angular equations are of the confluent Heun type and the separation constant $A$ becomes a non trivial function of $a\omega$.

\subsection{Angular equation}

The gravity/CFT dictionary is obtained by setting $s=-2$ in (\ref{dicang})
\bea
\chi&=&{2\over z} - 1 \qquad, \qquad x_\chi= 4 a \omega\qquad,\qquad{}_{-2} S_{\ell m} = {\Psi(z) \over \sqrt{1-\chi^2}}\nn\\
 c^\chi&=& -2 ~,~ p_0^\chi= \frac{-2-m}{2}~,~ k^\chi_0=\frac{-2+m}{2} ~,~ A= u^\chi-\ft{9}{4}  +2 a  \omega (1-m)  - a^2 \omega ^2
\label{dicang1}
\eea
 The separation constant is determined by the quantization condition
\be
\mathfrak{a}^\chi=\ell+\ft12
\ee
Plugging this into $u^\chi (\mathfrak{a})$ and using the dictionary (\ref{dicang}) one finds
for the separation constant
   \be
A=(\ell-1) (\ell+2)-\frac{8  m a \omega }{\ell (\ell+1)}+ a^2 \omega^2 \left[ H(\ell+1)-H(\ell)-1) \right]+\ldots
     \ee
   with
   \be
   H(\ell)={ 2(\ell^2-m^2)(\ell^2-4)^2 \over \ell^3(4\ell^2-1)}
   \ee
   This formula reproduces (92) of \cite{Sasaki:2003xr} after the identification $A=\lambda+2 m a \omega$.
 The eigenfunctions can be also obtained  iteratively order by order in $a \omega$ starting from the ansatz
 \bea
 {}_{-2}S_{\ell m}(\theta) &=& {}_{-2}Y_{\ell m}(\theta) + \sum_{n=1} \sum_{i=-n}^n  (a \omega )^n  d_{n\ell m, i} ~  {}_{-2}Y_{\ell+i, m}(\theta)   \eea
 with $ {}_{s}Y_{\ell m}$ given by (\ref{ylm}),  the solution at $a\omega=0$. At linear order in $a\omega$, one finds
 \bea
d_{1\ell m,1}  &=& \frac{2 }{(\ell {+}1 )^2}
 \sqrt{\frac{(\ell{-}1) (\ell{+}3) (1{+}\ell{-}m) (1{+}\ell{+}m)}{(1{+}2 \ell) (3{+}2
   \ell)}} \nn\\
  d_{1\ell m,-1} &=& -\frac{2}{\ell^2} \sqrt{\frac{(\ell-2) (\ell+2) (\ell-m) (\ell+m)}{(-1+2 \ell)
   (1+2 \ell)}}
 \eea
and $d_{1\ell m,0}=0$.  At order $(a \omega)^2$,   only the $\ell=2$ harmonics contribute till order $v^6$. The non-trivial coefficients are
\be
d_{22m,0}= {-}\ft12 \left(  d_{1\ell m,1}^2 {+}   d_{1\ell m,-1}^2 \right) ~ , ~  d_{22m,1}={m\sqrt{9{-}m^2}\over 324 \sqrt{7} }
 ~ , ~   d_{22m,2}={11\sqrt{(9{-}m^2)(16{-}m^2)}\over 1764 \sqrt{3} }
\ee

 \subsection{PN expansion of $R_{in}$ and $B_{in}$}

  The solution of the homogeneous radial equation with incoming boundary conditions and the asymptotic coefficient are given again by the CFT formula
 \bea
 R_{\rm in} &=&  z^{-2} (1-z)^{1\over 2}  \sum_{\alpha = \pm} F^{-1}_{- \alpha }  \Psi_\alpha(z)\nn\\
B_{\rm in} &=& 2\delta \, C_- = 2\delta \sum_{\alpha}     F^{-1}_{- \alpha}
B^{\rm conf}_{\alpha  -}   x^{{-}{1\over 2}{+}c{+}\alpha \mathfrak{a}}  e^{{ 1\over 2}  ( \partial_c-\alpha \partial_{\mathfrak{a}}) {\cal F}_{\rm inst}    }   \label{binexp2}
\eea
  with  radial dictionary
  \bea
 z &=& {2\delta \over r-M+\delta}   \qquad, \qquad  R_{\rm in} = (1-z)^{1\over 2} \, z^{2}\,  \Psi(z)  \nn\\
 x &=& 4 i \delta  \omega  \qquad , \qquad c= {-}2{-}2 i M \omega \qquad , \qquad k_0 =
 -1- iP_+ \qquad , \qquad  p_0= 1 +iP_- \nn\\
     u & =& A{+}\ft94{+}2 \omega ( a m{-} i \delta ){+}  \omega ^2(a^2{-}12 M^2{-} 8 M \delta) \label{dicr2}
     \eea
    with
    \bea
    r_{\pm}&=&M\pm \delta=M\pm \sqrt{M^2-a^2}\qquad , \qquad P_\pm=\frac{   a m{-}2  M \omega  r_\pm}{r_+-r_- }
    \eea

 The result to order $v^4$ is
 {\small
  \bea
  &&   \mathfrak{R}(r) = {R_{\rm in} \over 2{\rm i} \omega B_{\rm in} } =\frac{  (2\delta)^{-1-c}  r^{\mathfrak{a}+\frac{3}{2}} \left( -2 {\rm i} \omega  \right)^{\mathfrak{a}-\frac{1}{2}-c} \Gamma \left(c{+}\mathfrak{a}{+}\frac{1}{2}\right)}{ \Gamma \left(2 \mathfrak{a}+1\right)}
\left[
  1+\frac{4 i r  \omega }{1+2 \mathfrak{a}}
   -\frac{r^2 \omega ^2 \left(17+2 \mathfrak{a}\right)}{4 \left(1+\mathfrak{a}\right) \left(1+2
   \mathfrak{a}\right)} \right. \nn\\
&&
 -\frac{8 i a m-3 \delta +4 \delta  \mathfrak{a}+4 \delta  \mathfrak{a}^2}{2 r \left(-1+2
   \mathfrak{a}\right)} -\frac{4 i M \omega }{1+2 \mathfrak{a}}-\frac{i r^3
   \omega ^3 \left(7+2 \mathfrak{a}\right)}{\left(1+\mathfrak{a}\right) \left(1+2 \mathfrak{a}\right) \left(3+2
   \mathfrak{a}\right)}
  \label{rkerr} \\
&&
+\frac{\omega  \left(7+6 \mathfrak{a}\right) \left(8 a m+i \delta -4 i \delta
   \mathfrak{a}^2\right)}{\left(-1+2 \mathfrak{a}\right) \left(1+2 \mathfrak{a}\right){}^2} +\frac{r^4 \omega ^4 \left(163+72 \mathfrak{a}+4 \mathfrak{a}^2\right)}{32 \left(1+\mathfrak{a}\right)
   \left(2+\mathfrak{a}\right) \left(1+2 \mathfrak{a}\right) \left(3+2 \mathfrak{a}\right)}
\nn\\
&& +  r \omega ^2 \left(\frac{i a m \left(117+76 \mathfrak{a}+4 \mathfrak{a}^2\right)}{\left(1+\mathfrak{a}\right)
   \left(-1+2 \mathfrak{a}\right) \left(1+2 \mathfrak{a}\right){}^2}+\frac{12 M \left(3-2 \mathfrak{a}\right)+\delta
   \left(83+72 \mathfrak{a}+4 \mathfrak{a}^2\right)}{8 \left(1+\mathfrak{a}\right) \left(1+2 \mathfrak{a}\right)}\right)    \nn\\
&& \left.  \frac{1}{r^2} \left( \frac{a^2 m^2 \left(2 \mathfrak{a}{-}17 \right)}{4 \left(\mathfrak{a}{-}1\right) \left(2
   \mathfrak{a}{-}1 \right)}{-}\frac{1}{2} M \delta  \left(3{+}2 \mathfrak{a}\right){+}\frac{\delta ^2 \left(3{+}2 \mathfrak{a}\right)
   \left(4 \mathfrak{a}{+}4 \mathfrak{a}^2{-}11\right)}{16 \left(-1+\mathfrak{a}\right)}{+}\frac{2 i a m \left(\delta
   \left(3{+}2 \mathfrak{a}\right){-}2 M\right)}{2 \mathfrak{a}{-}1}\right)  +\ldots   \right]\nn
   \eea
   }
with $\mathfrak{a}$ given again by\footnote{In Appendix \ref{aExpansion} we give this quantity up to $(2M\omega)^4$.}
\be
\mathfrak{a}=\ell+\frac{1}{2} -(2\omega M)^2\frac{\left(15 \ell^4+30 \ell^3+28 \ell^2+13 \ell+24\right) }{2 \ell (\ell+1) (2 \ell-1) (2 \ell+1) (2 \ell+3)}+\ldots \label{p2final2}
\ee
Plugging (\ref{p2final2})  into (\ref{rkerr}) one finds
\bea
&& \mathfrak{R}(r) = -\frac{2^{\ell+2}  \Gamma (\ell-1) (-i r \omega )^{\ell+2}}{\Gamma (2\ell +2)}  \left[  1+\frac{2 i r \omega }{\ell+1}-\frac{2 i a m}{\ell r}-\frac{(\ell+2)
   M}{r}-\frac{( \ell+9) r^2 \omega ^2}{2 (\ell+1) (2 \ell+3)} \right. \nn\\
&&    \left(\frac{2 a
   (3\ell+5) m}{\ell (\ell+1)^2}-{\rm i} \delta  -\frac{2 {\rm i} (\ell+3) M  }{1+l} \right) \omega -\frac{ {\rm i} (\ell+4) r^3 \omega ^3}{(\ell+1) (\ell+2)
   (2\ell+3)}  \label{rlkerr} \\
 &&     +M \omega \left(\pi +2 {\rm i} \log \left(4 v^3 \delta  \omega \right)-2 {\rm i} \psi
   _0(\ell-1)\right)+\frac{\left(50+19 l+l^2\right) r^4 \omega ^4}{8 (1+l) (2+l) (3+2 l) (5+2 l)}\nn\\
&& {1\over r^2} \left( \frac{a^2 (-8+l) m^2}{2 l (-1+2 l)}+\frac{2 i a (1+l) m M}{l}+\frac{1}{2} (1+l) (2+l)
   M^2-\frac{(1+l) (2+l) \delta ^2}{2 (-1+2 l)} \right) \nn\\
   && \left. \frac{i a \left(39+20 l+l^2\right) m}{l (1+l)^2 (3+2 l)}+\frac{\left(24+5 l+l^2\right)
   M}{2 (1+l) (3+2 l)}+\frac{2 i M \pi +2 \delta +4 M \left(-\log \left(4 v^3 \delta  \omega
   \right)+\psi _0(-1+l)\right)}{1+l}   +\ldots
   \right] \nn
\eea

\subsection{Gravitational luminosity}

 Plugging (\ref{rlkerr}) into (\ref{dedt}) one finds for the gravitational luminosity
\bea
&& {dE\over dt} = {32  \mu^2 v^{10} \over 5 M^2} \left[
1-\frac{1247 v^2}{336}+\left(4 \pi -\frac{73 \mathfrak{q}}{12}\right)
   v^3+\left(-\frac{44711}{9072}+\frac{33 \mathfrak{q}^2}{16}\right) v^4+\frac{1}{672} (-8191 \pi
   +7498 \mathfrak{q}) v^5
 \right. \nn\\
&&  \left.+ v^6 \left(\frac{6643739519}{69854400}-\frac{1712 \gamma }{105}+\frac{16
   \pi ^2}{3}-\frac{169 \pi  \mathfrak{q}}{6}+\frac{3419 \mathfrak{q}^2}{168}-\frac{3424 \log
   (2)}{105}-\frac{1712 \log (v)}{105}\right)+\ldots \right]
\eea
in agreement with  previous results in the literature \cite{Shibata:1994jx,Tanaka:1996ht,Mino:1997bx,Sasaki:1981sx,Sasaki:2003xr}.

  \section{ Tidal interactions}

  Tidal interactions describe the response of a gravitational object to small perturbations of the geometry. They show up in the PN expansion of the wave form at order $v^{10}$, where the solutions $\Psi_\pm$ describing the propagation of the wave in the near zone start to mix.
 The strength of the tidal interactions is measured by the Love and dissipation numbers defined as the real and imaginary parts of the ratio
 between the coefficients of the source and response components $\Psi_\pm$ of the solution  in the near zone.  This ratio is sensitive to the boundary conditions defining $\Psi(z)$ and provide therefore a quantitative measure of the compactness properties of the geometry.

 In this section, following \cite{Consoli:2022eey}, we compute the dynamical Love and dissipation numbers for Kerr BH's and Kerr like compact objects. A spin s perturbation of a Kerr BH is described by a solution of the confluent Heun equation with incoming boundary conditions $\Psi=\widehat{\Psi}_-$ leading to
   \bea
  R_{\rm Kerr} (z)   &=&     (1-z)^{-{(s+1)\over 2}} \, z^s\,  \sum_{\alpha=\pm} F^{-1}_{- \alpha}  \Psi_\alpha(z)  \label{rhom}
\eea
   with
   \be
   \Psi_\alpha(z)  \underset{z,\ft{\omega}{z}\to 0}{\approx} z^{{1\over 2}+\mathfrak{a}} (1+\ldots )
   \ee
   The dynamical Love and dissipation numbers are given as the real and imaginary parts of the ratio of the $\Psi_\pm$ coefficients
     \bea
{\cal L}_{\rm Kerr} &=&  {F^{-1}_{-+} \over F^{-1}_{--} }=
\frac{\Gamma (-2 \mathfrak{a} ) \Gamma (\ft12+m_1+\mathfrak{a}  ) \Gamma (\ft12+m_2+\mathfrak{a}  )}{
\Gamma (2 \mathfrak{a} ) \Gamma (\ft12+m_1-\mathfrak{a}  ) \Gamma (\ft12+m_2-\mathfrak{a}  ) }
\eea
with
 \bea
 m_1 &=& k_0+p_0=s+ {\rm i} \epsilon  \quad, \quad  m_2=k_0-p_0=-2 {\rm i} P_+   -{\rm i} \epsilon \nn\\
 \mathfrak{a} &=& \widehat{\ell}  +\ft12 =  \ell +\ft{1}{2}-\epsilon^2\frac{\left(15 \ell^4+30 \ell^3+28 \ell^2+13 \ell+24\right) }{2 \ell (\ell+1) (2 \ell-1) (2 \ell+1) (2 \ell+3)}+ O(\epsilon^4)
      \eea
Here we introduced the notion of ``quantum" orbital number $\widehat\ell(\omega)$, that  account for quantum corrections of the SW period
$\mathfrak{a}(\epsilon)$.  The non-trivial $\omega$-dependence of the tidal response function ${\cal L}_{\rm Kerr}$ is encoded in the replacement $\ell \to \widehat{\ell}$.

  The analysis above can be easily adapted to the case of a compact Kerr like geometry. This geometry is obtained
   by cutting out a ball of radius $r_+ + \Delta r$ around the origin and replacing
  the internal region by a horizonless geometry. Since the metric in the external region is that of Kerr, the solution outside is given again by a linear combination of $\Psi_\alpha$. The precise combination depends on the reflectivity assumed at the boundary of the two regions.
  For simplicity, we assume here that  $\Delta r \ll  r_+$, so boundary conditions are imposed at $r=r_+$, where now a partial reflection of the wave is allowed.  We write
  \be
  \Psi=\sum_{\alpha''} D_{\alpha''}\widehat{\Psi}_{\alpha''} (z) =\sum_{\alpha,\alpha''} D_{\alpha''} F^{-1}_{\alpha'' \alpha }  \Psi_\alpha(z) \label{psiright}
  \ee
  with
     \be
     {\cal R}={D_+\over D_-}
    \ee
  a number, parametrizing the reflectivity of the boundary.   Expanding the right hand side of (\ref{psiright}) in the near zone, one finds
    \be
  R_{\rm ECO} (z)   \sim \Psi_-(z)+ {\cal L}_{\rm ECO} \Psi_+(z)\sim
  \left[ z^{ s-\widehat{\ell} }  (1+\ldots)  + z^{s+\widehat{\ell}+1  } \,{\cal L}_{\rm ECO}  (1+ \ldots)   \right]  \label{rintheta0}
 \ee
 with
 \bea
{\cal L}_{ECO}= { \sum_{\alpha''} D_{\alpha''} F^{-1}_{\alpha'' +}   \over \sum_{\alpha''} D_{\alpha''} F^{-1}_{\alpha'' -} } = { {\cal L}_{\rm Kerr} + {\cal R}  \gamma_{+}\over 1+ {\cal R}  \gamma_{-} } \label{leco}
\eea
 and
   \bea
\gamma_{+}  &=&  {F^{-1}_{+ + } \over F^{-1}_{--} }=
\frac{\Gamma ({-}2\widehat{\ell} {-}1  ) \Gamma (\widehat{\ell}{+}1{+}s {+} i  \epsilon )  \Gamma ( 1{-}s{+} 2 {\rm i} P_+ ) \Gamma (\widehat{\ell}{+}1{-}2 {\rm i} P_+{-} {\rm i}  \epsilon )}{\Gamma (2 \widehat{\ell}{+}1)
    \Gamma \left(  {-}  \widehat{\ell} {-}s  {-} {\rm i} \epsilon \right)  \Gamma \left(1{+} s {-} 2 {\rm i}P_+ \right) \Gamma ( {-}\widehat{\ell} {+}2 {\rm i} P_+
 {+} i  \epsilon) } \nn\\
 \gamma_{-}  &=&  {F^{-1}_{+ - } \over F^{-1}_{--} }=
\frac{ \Gamma (\widehat{\ell}{+}1{+}s {+} i  \epsilon )  \Gamma ( 1{-}s{+} 2 {\rm i} P_+ ) \Gamma (\widehat{\ell}{+}1{-}2 {\rm i} P_+{-} {\rm i}  \epsilon )}{  \Gamma \left(    \widehat{\ell} {+}1{-}s  {-} {\rm i}  \epsilon \right)  \Gamma \left(1{+} s {-} 2 {\rm i}P_+ \right) \Gamma ( \widehat{\ell}{+}1 {+}2 {\rm i} P_+
 {+} i  \epsilon) }
\eea
 In the static limit $\omega \to 0$ one finds
 \bea
{\cal L}_{\rm Kerr} &=&  {\rm i} P_+ (-)^{s-\ell-1}
\frac{   ( \ell-s )! (\ell+s)!}{ (2\ell)! (2\ell+1)^2  } \prod_{n=1}   \left( n^2 +4 P_{+}^2 \right)
     \label{love4}\\
\gamma_{+} &=&
 {\rm i} P_{+}(-)^{s-\ell-1}  \frac{   (\ell{-}s )!(\ell+s)! \Gamma (1{+}s{-}2 {\rm i} P_{+} )}{ \ell! (2\ell+1)!
     \Gamma ( 1 {-}s{+}2 {\rm i} P_{+}
 ) } \prod_{n=1}   \left( n^2 +4 P_{+}^2 \right) \nn\\
 \gamma_{-} &=&
\frac{ (2\ell)! (\ell{-}s )! \Gamma (1{+}\ell{+}2 {\rm i} P_{+}  ) \Gamma (1{+}s{-}2 {\rm i} P_{+}  )}{ \ell! (\ell {+}s)!   \Gamma \left(1  {+} \ell{-}2 {\rm i}P_{+}  \right)   \Gamma ( 1 {-}s{+}2 {\rm i} P_{+}
 ) }\nn\\
\eea

\section{Summary and Discussion}

In this paper we derive a combinatorial formula for the PN expansion of the wave form produced by particles orbiting around Schwarzschild and Kerr BH's. To this aim, we exploit techniques coming from Seiberg-Witten theory, localization and AGT duality. In this framework the wave form is written as the partition function of an $SU(2)^2$ quiver gauge theory given as a double series with expansion parameters
accounting for, on the gravity side, to Post Newtonian and Post Minkowskian corrections. All log's and trascendental numbers typically appearing in the PN expansions are resummed into exponential and Gammma functions. The results are in agreeement with those appeared in the literature  obtained with other computational techniques and provide a compact formula for the wave form valid for arbitrary values of the orbital quantum number $\ell$.

The results show that, in the PN limit, the wave form is dominated by one of the two solutions of the Heun equation describing the gravitational perturbations at linear order. The second solution starts to contribute at the 5-th PN order beyond the quadrupole approximation. This mixing is encoded in the tidal Love function carrying the information of the boundary conditions specifying the gravitational object. We derive a formula for the tidal Love function for Kerr BH's and Kerr like ECO's.

It would be interesting to apply the techniques in this paper to the case of open orbits for which the Post Newtonian and the Post Minkoskian approximations are not directly related as in the case of circular orbits. This line of research would certainly contribute to the recent efforts to relate the results coming from scattering amplitudes, effective theories, worldline quantum field theories etc.
Each one of these techniques performs better than others in particular regions
providing complementary viewpoints which can be combined to improve our actual understanding.

\section*{Acknowledgements}
We acknowledge fruitful scientific exchange with M. Bianchi, D. Bini,  G.~Bonelli, G. Di Russo,  P. Pani,  R. Russo and A.~Tanzini. We thank C. Iossa for discussions and collaborations at the early stages of this work.
 We thank the MIUR PRIN contract 2020KR4KN2 ``String Theory as a bridge between Gauge Theories and Quantum Gravity'' and the INFN project ST\&FI ``String Theory and Fundamental Interactions'' for partial support.

\begin{appendix}

	\section{CFT tools}\label{cfttools}

	\subsection{Conformal blocks}\label{CFTconfblocks}
	
	We consider a 2d Liouville theory with background charge
	\be
	Q=b+{1\over b}
	\ee
	We consider an $(n+3)$-point correlator  of primary fields, with vertices $V_{P_s}=e^{(Q-2P_s) \phi}$, momenta
	\be
	P_s=\ft{1}{b}(p_0,k_0,k_1\ldots k_n,p_{n+1})
	\ee
	and dimensions
	\be
	\Delta_{p_i}={Q^2\over 4} -{p_i^2 \over b^2} \qquad , \qquad   \Delta_{k_i}={Q^2\over 4} -{k_i^2 \over b^2}
	\ee
	The $(n+3)$-point conformal block will be denoted by\footnote{
		As usual, we define $\langle   p_0 |   =\underset{z_{-1} \to \infty}{\lim}   \langle  0 | z_{-1}^{2\Delta_{p_0} } V_{p_0} (z_{-1} ) $ and $| p_{n+1}\rangle= V_{p_{n+1}}(0 ) |0\rangle $  }
	\be
	\cF_{p_0}{}^{k_0}{}_{p_1} .. {}^{k_{n}}{}_{p_{n+1} }   (z_i)
	=     \left \langle   p_0 |  V_{k_0}(z_0)  \ldots V_{k_{n} }(z_{n} )  | p_{n+1}  \right\rangle _{p_1\ldots p_n} =
	\begin{tikzpicture}[scale=0.8,baseline={([yshift=-.5ex]current bounding box.center)}]
		\draw [] (11,0)--(13.2,0);
		\draw [] (13.4,0)--(13.5,0);
		\draw [] (13.6,0)--(14.1,0);
		\draw [] (14.1,0)--(15.1,0);
		\node [left] at (11,0) {\scriptsize{$\infty$}};
		\node [below] at (11.6,0) {\scriptsize{$ p_0$}};
		\node [below] at (12.6,0) {\scriptsize{$p_{1}$}};
		\node [below] at (14,0) {\scriptsize{$p_{n}$}};
		\node [below] at (14.95,0) {\scriptsize{$ p_{n+1}$}};
		\node [right] at (15.1,0) {\scriptsize{$0$}};
		\node [left] at (12,0.7) {\scriptsize{$z_0$}};
		\node [above] at (12,1) {\scriptsize{$k_0$}};
		\draw [] (12,0)--(12,1);
		\draw [] (13,0)--(13,1);
		\node [left] at (13,0.7) {\scriptsize{$z_1$}};
		\node [above] at (13,1) {\scriptsize{$k_1$}};
		\draw [] (14.3,0)--(14.3,1);
		\node [left] at (14.3,0.7) {\scriptsize{$z_n$}};
		\node [above] at (14.3,1) {\scriptsize{$k_n$}};
	\end{tikzpicture}
	\ee
	The parameters  $p_1, p_2, \ldots p_n$ specify the intermediate momenta.
	According to the AGT correspondence the conformal blocks of the Liouville theory can be related to the instanton partition function of a quiver gauge theory with gauge group $SU(2)^n$ written as
	\be
	\mathcal{F}_{p_0}{}^{k_0}.. {}^{k_{n}}{}_{p_{n+1} }   (z_i) =  \prod_{i=0}^n   z_{i}^{\Delta_{p_i} {-}\Delta_{p_{i+1} }{-}\Delta_{k_i}}    \prod_{ \underset{i< j}{i,j=0} }^{2}  \left(1{-}{z_{j} \over z_i} \right)^{ (2k_{i}{-}b Q)(2k_{j}{+}b Q) \over 2 b^2  }  \,
	Z_{\rm inst}{}_{p_0}{}^{k_0} .. {}^{k_{n}}{}_{p_{n+1} }     \label{ffinst}
	\ee
	with $Z_{\rm inst}{}_{p_0}{}^{k_0} .. {}^{k_{n}}{}_{p_{n+1} }  (q_i)$ given in (\ref{inst1}),(\ref{inst2}), (\ref{inst3}).
Similarly, correlators with the insertion of a null field ot type $(2,1)$ will be represented as
\be
\Phi^{n+3}_\alpha(z,z_i) = \cF_{p_0}{}^{k_0}{}_{p_1} ....{}_{p_s^{-\alpha}}{}^{k_{\rm deg}} {}_{p_s}....  {}^{k_{n}}{}_{p_{n+1} }   (z_i)
=
\begin{tikzpicture}[scale=0.8,baseline={([yshift=-.5ex]current bounding box.center)}]
	\draw [] (11,0)--(13.2,0);
	\draw [] (13.4,0)--(13.5,0);
	\draw [] (13.6,0)--(14.1,0);
	\draw [] (14.1,0)--(15.1,0);
	\node [left] at (11,0) {\scriptsize{$\infty$}};
	\node [below] at (11.6,0) {\scriptsize{$ p_0$}};
	\node [below] at (12.6,0) {\scriptsize{$p_{1}$}};
	\node [right] at (17.3,0) {\scriptsize{$0$}};
	\node [left] at (12,0.7) {\scriptsize{$z_0$}};
	\node [above] at (12,1) {\scriptsize{$k_0$}};
	\draw [] (12,0)--(12,1);
	\draw [] (13,0)--(13,1);
	\node [left] at (13,0.7) {\scriptsize{$z_1$}};
	\node [above] at (13,1) {\scriptsize{$k_1$}};
	\draw [line,dashed] (14.3,0)--(14.3,1);
	\draw [line,dashed] (14.3,0)--(14.8,0);
	\node [left] at (15.6,0.7) {\scriptsize{$z_s=z$ }};
	\node [above] at (14.3,1) {\scriptsize{$k_{\rm deg}$}};
	\node [below] at (14,0) {\scriptsize{$p_{s}^{-\alpha}$}};
	\node [below] at (14.8,0) {\scriptsize{$ p_{s}$}};
	\draw [line,dashed] (15.45,0)--(16,0);
	\draw [] (16.3,0)--(16.3,1);
	\node [left] at (17,0.7) {\scriptsize{$z_n$}};
	\node [above] at (16.3,1) {\scriptsize{$k_n$}};
	\node [below] at (16,0) {\scriptsize{$p_{n}$}};
	\node [below] at (16.8,0) {\scriptsize{$ p_{n+1}$}};
	\draw [] (15.8,0)--(17,0);
\end{tikzpicture}
\ee
   with
	\bea
	k_{\rm deg}=\ft{1}{2} + b^2 \qquad \Delta_{\rm deg}={-}\ft12{-}\ft{3 b^2}{4}
	\label{vnull1}
	\eea
	 and
	\be
	p_s^\alpha = p_s+\frac{\alpha b^2}{2}
	\ee
	with $\alpha=\pm$.
	
	\subsection{The confluent limit}\label{CFTconfluent}
	
	The degenerated five point conformal blocks
	\be
 	\begin{tikzpicture}[baseline={(current bounding box.center)}, node distance=0.8cm and 0.8cm]
		\coordinate[label=above:$k_0$] (k0);
		\coordinate[below=of k0] (s0);
		\coordinate[left=of s0,label=left:\scriptsize{$\infty$}] (p0);
		\coordinate[right=1cm of s0] (s1);
		\coordinate[above=of s1,label=above:$k_{\rm deg}$] (k1);
		\coordinate[right=of s1] (p2);
		\coordinate[right=1cm of s1] (s2);
		\coordinate[above=of s2,label=above:$k_2$] (k2);
		\coordinate[right=of s2, ,label=right:\scriptsize{$0$}] (p3);
		\draw[line] (k0) -- (s0);
		\draw[line,dashed] (s0) --node[label={[xshift=0.2cm, yshift=0.1cm]left:\scriptsize{$ 1$}}] {}  (k0);
		\draw[line] (s0) -- (p0);
		\draw[line,dashed] (k1) -- node[label={[xshift=0.2cm, yshift=0.1cm]left:\scriptsize{$ z$}}] {} (s1);
		\draw[line] (s1) -- (p2);
		\draw[line] (s0) -- node[label={[yshift=0.2cm]below:$\mathfrak{a}^{-\alpha}$}] {} (s1);
		\draw[line] (s0) -- node[label=below:$p_0$] {} (p0);
		\draw[line] (k2) -- node[label={[xshift=0.2cm, yshift=0.1cm]left:\scriptsize{$ z_2$}}] {} (s2);
		\draw[line] (s2) -- (p3);
		\draw[line] (s1) -- node[label=below:$\mathfrak{a}$] {} (s2);
		\draw[line] (s2) -- node[label=below:$p_3$] {} (p3);
	\end{tikzpicture}
	\ee
	satisfy a differential equation of the Heun type  in the variable $z$. The confluent Heun equation is obtained by considering the
	 limit where two singularities collide at $z=0$.  More precisely, we take  $z_2$ to zero, $k_2$, $p_3$ to infinity, keeping finite the combinations
	\be
	c=k_2+p_3 \qquad , \qquad  x= z_2 (k_2-p_3). \label{conflim}
	\ee
	The confluent limits of the relevant four and five point conformal block obtained will be denoted as
		\bea
	\Phi^5_\alpha (z)  &=&  \begin{tikzpicture}[baseline={(current bounding box.center)}, node distance=1.0cm and 1.0cm]
		\coordinate[label=right:$k_0$] (k0);
		\coordinate[below=of k0] (s0);
		\coordinate[left=of s0,label=left:\scriptsize{$\infty$}] (p0);
		\coordinate[right=of s0] (s1);
		\coordinate[right=of s1] (s2);
		\coordinate[right=of s2] (s3);
		\coordinate[above=of s1,label=right:$k_{\rm deg}$] (k);
		\draw[line,dashed] (s1) --node[label=right:$z$] {} (k);
		\draw[line] (p0) --node[label={[yshift=0.0cm]below:$ p_0$}] {}  (s0);
		\draw[line] (s0) --node[label={[yshift=0.2cm]below:$ \mathfrak{a}^{-\alpha}$}] {}  (s1);
		\draw[line] (s1) -- node[label=below:$\mathfrak{a}$] {} (s2);
		\draw[line,double] (s2) -- node[label=below:$c$] {} (s3);
		\draw[line] (s0) --node[label=right:\scriptsize{$ 1$}] {} (k0);
		\filldraw (s2) circle (3pt) node[above=2pt] {$x$ };
	\end{tikzpicture}
	\equiv
	\lim_{\begin{smallmatrix}   z_2 \to 0 \\
			k_2,p_3 \to \infty   \end{smallmatrix}}
	\begin{tikzpicture}[baseline={(current bounding box.center)}, node distance=0.8cm and 0.8cm]
		\coordinate[label=above:$k_0$] (k0);
		\coordinate[below=of k0] (s0);
		\coordinate[left=of s0] (p0);
		\coordinate[right=1cm of s0] (s1);
		\coordinate[above=of s1,label=above:$k_{\rm deg}$] (k1);
		\coordinate[right=of s1] (p2);
		\coordinate[right=1cm of s1] (s2);
		\coordinate[above=of s2,label=above:$k_2$] (k2);
		\coordinate[right=of s2] (p3);
		\draw[line] (k0) -- (s0);
		\draw[line,dashed] (s0) --node[label={[xshift=0.2cm, yshift=0.1cm]left:\scriptsize{$ 1$}}] {}  (k0);
		\draw[line] (s0) -- (p0);
		\draw[line,dashed] (k1) -- node[label={[xshift=0.2cm, yshift=0.1cm]left:\scriptsize{$ z$}}] {} (s1);
		\draw[line] (s1) -- (p2);
		\draw[line] (s0) -- node[label={[yshift=0.2cm]below:$\mathfrak{a}^{-\alpha}$}] {} (s1);
		\draw[line] (s0) -- node[label=below:$p_0$] {} (p0);
		\draw[line] (k2) -- node[label={[xshift=0.2cm, yshift=0.1cm]left:\scriptsize{$ z_2$}}] {} (s2);
		\draw[line] (s2) -- (p3);
		\draw[line] (s1) -- node[label=below:$\mathfrak{a}$] {} (s2);
		\draw[line] (s2) -- node[label=below:$p_3$] {} (p3);
	\end{tikzpicture} \nn\\
	\Phi_4 &=&
	\begin{tikzpicture}[baseline={(current bounding box.center)}, node distance=1.0cm and 1.0cm]
		\coordinate[label=above:$$] (s0);
		\coordinate[left=of s0] (p0);
		\coordinate[left=of p0] (p00);
		\coordinate[above=of p0,label=right: $k_0$] (k0);
		\draw[line] (p0) --node[label=right:\scriptsize{$ 1$}] {} (k0);
		\draw[line] (p0) --node[label=below:$ p_0$] {}  (p00);
		\coordinate[right=1cm of s0] (s1);
		\draw[line] (s0) --node[label=below:$\mathfrak{a}$] {} (p0);
		\draw[line,double] (s0) -- node[label=below:$c$] {} (s1);
		\filldraw (s0) circle (3pt) node[above=2pt] {${x}$ };
	\end{tikzpicture}
	\equiv  \lim_{\begin{smallmatrix}  z_2 \to 0 \\ k_2,p_3 \to \infty   \end{smallmatrix}}
	\begin{tikzpicture}[baseline={(current bounding box.center)}, node distance=0.8cm and 0.8cm]
		\coordinate[label=above:$k_{0}$] (k0);
		\coordinate[below=of k0] (s0);
		\coordinate[left=of s0] (p0);
		\coordinate[right=1cm of s0] (s1);
		\coordinate[above=of s1,label=above:$k_2$] (k1);
		\coordinate[right=of s1] (p2);
		\draw[line] (k0) -- node[label={[xshift=0.2cm, yshift=0.1cm]left:\scriptsize{$ 1$}} ] {} (s0);
		\draw[line] (s0) -- (p0);
		\draw[line] (s1) --node[label={[xshift=0.2cm, yshift=0.1cm]left:\scriptsize{$ z_2$}}] {}  (k1);
		\draw[line] (s1) -- (p2);
		\draw[line] (s0) -- node[label=below:$\mathfrak{a}$] {} (s1);
		\draw[line] (s0) -- node[label=below:$p_0$] {} (p0);
		\draw[line] (s1) -- node[label=below:$p_3$] {} (p2);
	\end{tikzpicture}
	\eea
In this limit the last contribution to the instanton partition function in  (\ref{inst2}) is replaced by
\be
z_2^{|Y_2|}  z^{\rm bifund}_{Y_{2},\emptyset }(   \mathfrak{a} , p_3 ,- k_2 ) \to  x^{|Y_2|} z^{\rm hyp}_{Y_{2} }(   \mathfrak{a} ,-c )
\ee	
with
\be
z^{\rm hyp}_{Y_{2} }(   \mathfrak{a} ,-c )= \prod_{\alpha}
	    \prod_{s\in Y_\alpha}  \left[  -E_{Y_\alpha,\emptyset}(\alpha \mathfrak{a}{+} c,s) {+} \ft{\epsilon}{2}   \right]
\ee	
On the other hand in the confluent limit
\be
 \left(1{-}{z_{2} \over z} \right)^{ (2k_{\rm deg}{-}b Q)(2k_{2}{+}b Q) \over 2 b^2  } \to  e^{ {x\over 2 z} }
\ee
We are interested in the $b \to 0$ limit. It is easy to see that  both  $\Phi^5_\alpha$ and $\Phi_4$  are divergent but  their ratio is finite
	\be
	\Psi_\alpha(z)  =  \lim_{b \to 0} {\Phi^5_\alpha(z) \over \Phi_4 }  =e^{{x\over 2z} }  \,  z^{{1\over 2} +\alpha \mathfrak{a}}     \,   \lim_{b \to 0}
\left(1{-} z  \right)^{ (2k_{0}{-}bQ )( 2 k_{\rm deg} {+}  b Q ) \over 2 b^2  }   {
	Z_{\rm inst}{}_{p_0}{}^{k_0}  {}_{\mathfrak{a}^{-\alpha}}{}^{ k_{\rm deg}}  {}_{\mathfrak{a}}  {}_c  (1,z,x) \over
	Z_{\rm inst}{}_{p_0}{}^{k_0} {}_{\mathfrak{a}}   {}_c (1,x)   }
	  \, \label{psidef1}
	\ee
	The first few terms in the instanton expansion are given in (\ref{psialpha}). 	 Similarly, the description of the solution in the near horizon zone can be obtained from $\Psi_\alpha(z) $ with the use of braiding and fusion relations. Diagramatically
	\be
	\begin{array}{ccccc}
		{\rm Near~ horizon} (z\approx 1) & &  {\rm Near~ zone} (1\gg z \gg x)&&  {\rm Far ~away}( 1 \gg x  \gg z ) \\
		\begin{tikzpicture}[scale=0.5,baseline={(current bounding box.center)}, node distance=0.6cm and 0.6cm]
			\coordinate[label=left:$k_0$] (k0);
			\coordinate[below right=of k0] (s0);
			\coordinate[above right=of s0,label=right:$k_{\rm deg}$] (k1);
			\coordinate[below=of s0] (s1);
			\coordinate[left=of s1] (p0);
			\coordinate[right=of s1,xshift=0.2cm] (s2);
			\draw[line] (k0) -- node[label={[xshift=-0.2cm, yshift=0.2cm]below:\scriptsize{$ x$}}] {} (s0);
			\draw[line,dashed] (s0) -- node[label={[xshift=0.2cm, yshift=0.2cm]below:\scriptsize{$ z$}}] {} (k1);
			\draw[line] (s0) -- node[label={[xshift=-0.2cm]right:\scriptsize{$ k_{0\alpha''}$}}] {} (s1);
			\draw[line] (s1) -- node[label=below:$p_0$] {} (p0);
			\draw[line] (s1) -- node[label=below:$\mathfrak{a}$] {} (s2);
			\coordinate[right=of s2] (s3);
			\draw[line,double] (s2) -- node[label=below:$c$] {} (s3);
			\filldraw (s2) circle (5pt) node[above=2pt] {${x}$ };
		\end{tikzpicture}
		&=& F^{-1}_{\alpha'' \alpha}
		\begin{tikzpicture}[baseline={(current bounding box.center)}, node distance=1cm and 1cm]
			\coordinate[label=right:$k_0$] (k0);
			\coordinate[below=of k0] (s0);
			\coordinate[left=of s0] (p0);
			\coordinate[right=of s0] (s1);
			\coordinate[right=of s1] (s2);
			\coordinate[right=of s2] (s3);
			\coordinate[above=of s1,label=right:$k_{\rm deg}$] (k);
			\draw[line,dashed] (s1) --node[label=right:$z$] {} (k);
			\draw[line] (p0) --node[label={[yshift=0.0cm]below:$ p_0$}] {}  (s0);
			\draw[line] (s0) --node[label={[yshift=0.2cm]below:$ \mathfrak{a}^{-\alpha}$}] {}  (s1);
			\draw[line] (s1) -- node[label=below:$\mathfrak{a}$] {} (s2);
			\draw[line,double] (s2) -- node[label=below:$c$] {} (s3);
			\draw[line] (s0) --node[label=right:\scriptsize{$ 1$}] {} (k0);
			\filldraw (s2) circle (3pt) node[above=2pt] {${x}$ };
		\end{tikzpicture}
		& =& F^{-1}_{\alpha'' \alpha}  B^{\rm conf}_{\alpha\alpha'}   \begin{tikzpicture}[baseline={(current bounding box.center)}, node distance=0.8cm and 0.8cm]
			\coordinate[label=right:$k_0$] (k0);
			\coordinate[below=of k0] (s0);
			\coordinate[left=of s0] (p0);
			\coordinate[right=of s0] (s1);
			\coordinate[right=of s1] (s2);
			\coordinate[right=of s2] (s3);
			\coordinate[above=of s2,label=right:$k_{\rm deg}$] (k);
			\draw[line,dashed] (s2) --node[label=right:$z$] {} (k);
			\draw[line] (p0) --node[label={[yshift=0.0cm]below:$ p_0$}] {}  (s0);
			\draw[line] (s0) --node[label={[yshift=0.2cm]below:$ \mathfrak{a}^{-\alpha}$}] {}  (s1);
			\draw[line,double] (s1) --node[label={[yshift=0.2cm]below:$ c^{-\alpha'}$}] {}  (s2);
			\coordinate[right=of s3] (s4);
			\draw[line,double] (s2) -- node[label=below:$c$] {} (s3);
			\draw[line] (s0) --node[label=right:\scriptsize{$ 1$}] {} (k0);
			\filldraw (s1) circle (3pt) node[above=2pt] {${x}$ };
		\end{tikzpicture}   \\
		\widehat{\Psi}_{\alpha''} (z)   & = &  F^{-1}_{\alpha \alpha'}  \Psi_\alpha(z)  &=& F^{-1}_{\alpha'' \alpha}  B^{\rm conf}_{\alpha\alpha'}   \widetilde{\Psi}_{\alpha'} (z)
	\end{array}
	\label{braidfus2}
	\ee
	with braiding and fusion matrices given by  (\ref{braidingBconf}).
	 In particular, the asymptotics at infinity is  given by \cite{Consoli:2022eey}
	\bea
	\Psi_{\rm in} (x,z)  &  = &\underset{z \to 0}{\lim_{b \to 0}}   \sum_{\alpha,\alpha'} F^{-1}_{- \alpha}
	B^{\rm conf}_{\alpha  \alpha'}   {     \begin{tikzpicture}[baseline={(current bounding box.center)}, node distance=0.8cm and 0.8cm]
			\coordinate[label=right:$k_0$] (k0);
			\coordinate[below=of k0] (s0);
			\coordinate[left=of s0] (p0);
			\coordinate[right=of s0] (s1);
			\coordinate[right=of s1] (s2);
			\coordinate[right=of s2] (s3);
			\coordinate[above=of s2,label=right:$k_{\rm deg}$] (k);
			\draw[line,dashed] (s2) --node[label=right:$z$] {} (k);
			\draw[line] (p0) --node[label={[yshift=0.0cm]below:$ p_0$}] {}  (s0);
			\draw[line] (s0) --node[label={[yshift=0.2cm]below:$ \mathfrak{a}^{-\alpha}$}] {}  (s1);
			\draw[line,double] (s1) --node[label={[yshift=0.2cm]below:$ c^{-\alpha'}$}] {}  (s2);
			\coordinate[right=of s3] (s4);
			\draw[line,double] (s2) -- node[label=below:$c$] {} (s3);
			\draw[line] (s0) --node[label=right:\scriptsize{$ 1$}] {} (k0);
			\filldraw (s1) circle (3pt) node[above=2pt] {${x}$ };
		\end{tikzpicture}  \over   \begin{tikzpicture}[baseline={(current bounding box.center)}, node distance=0.8cm and 0.8cm]
			\coordinate[label=above:$$] (s0);
			\coordinate[left=of s0] (p0);
			\coordinate[left=of p0] (p00);
			\coordinate[above=of p0,label=right: $k_0$] (k0);
			\draw[line] (p0) --node[label=right:\scriptsize{$ 1$}] {} (k0);
			\draw[line] (p0) --node[label=below:$ p_0$] {}  (p00);
			\coordinate[right=1cm of s0] (s1);
			\draw[line] (s0) --node[label=below:$\mathfrak{a}$] {} (p0);
			\draw[line,double] (s0) -- node[label=below:$c$] {} (s1);
			\filldraw (s0) circle (3pt) node[above=2pt] {${x}$ };
	\end{tikzpicture}     }=\nonumber\\
&= & \lim_{b \to 0} \sum_{\alpha,\alpha'} F^{-1}_{- \alpha}
B^{\rm conf}_{\alpha  \alpha'}{
	\begin{tikzpicture}[baseline={(current bounding box.center)}, node distance=0.8cm and 0.8cm]
		\coordinate[label=right:$k_0$] (k0);
		\coordinate[below=of k0] (s0);
		\coordinate[left=of s0] (p0);
		\coordinate[right=of s0] (s1);
		\coordinate[right=of s1] (s2);
		\coordinate[right=of s2] (s3);
         \coordinate[above=of s1,label=right:] (k);
		\draw[line] (p0) --node[label={[yshift=0.0cm]below:$ p_0$}] {}  (s0);
		\draw[line] (s0) --node[label={[yshift=0.2cm]below:$ \mathfrak{a}^{-\alpha}$}] {}  (s1);
		\draw[line,double] (s1) --node[label={[yshift=0.2cm]below:$ c^{-\alpha'}$}] {}  (s2);
		\draw[line] (s0) --node[label=right:\scriptsize{$ 1$}] {} (k0);
		\filldraw (s1) circle (3pt) node[above=2pt] {${x}$ };
\end{tikzpicture}  \over   \begin{tikzpicture}[baseline={(current bounding box.center)}, node distance=0.8cm and 0.8cm]
		\coordinate[label=above:$$] (s0);
		\coordinate[left=of s0] (p0);
		\coordinate[left=of p0] (p00);
		\coordinate[above=of p0,label=right: $k_0$] (k0);
		\draw[line] (p0) --node[label=right:\scriptsize{$ 1$}] {} (k0);
		\draw[line] (p0) --node[label=below:$ p_0$] {}  (p00);
		\coordinate[right=1cm of s0] (s1);
		\draw[line] (s0) --node[label=below:$\mathfrak{a}$] {} (p0);
		\draw[line,double] (s0) -- node[label=below:$c$] {} (s1);
		\filldraw (s0) circle (3pt) node[above=2pt] {${x}$ };
\end{tikzpicture}     }	\quad\	\bigtimes					
			\begin{tikzpicture}[baseline={(current bounding box.center)}, node distance=0.8cm and 0.8cm]
	\coordinate (s4);
	\coordinate[right=of s4] (s5);
	\coordinate[right=of s5] (s6);
	\coordinate[right=of s6] (s7);
	\coordinate[right=of s7] (s8);
	\draw[line,double] (s4) --node[label={[yshift=0.2cm]below:$  c^{-\alpha'}$}] {} (s5);
	\draw[line,double] (s5) --node[label={[yshift=-0.0cm]below:$ c$}] {} (s6);
	\coordinate[above=of s5,label=right:$k_{\rm deg}$] (kk);
	\draw[line,dashed] (s5) --node[label=right:$\ft{z}{x} $] {} (kk);
	\end{tikzpicture}		
				\label{connection1}
				\eea
 The three point function in (\ref{connection1}) can be computed by  expanding (\ref{hypf}) in the limit $z/x\ll 1$, while the ratio of four point functions  gives the instanton contribution. Finally comparing against (\ref{bcpsi2}) one finds (\ref{calphap}).

\section{The source term }\label{sourceterm}

 In this appendix we collect some formulae needed for the evaluation of the source term in the Teukolsky equations. We refer the reader to \cite{Mino:1997bx} and references therein, for details and explanations. We focus on spin $s=-2$ perturbations corresponding to components
  \be
 \Psi_{-2}=\rho^{-4} \psi_4 =-\rho^{-4} C_{\mu\nu\sigma \rho} n^\mu\overline{m}^\nu n^\sigma \overline{m}^\rho
 \ee
 with $\rho=(r-ia\cos\theta)^{-1}$,  $C_{\mu\nu\sigma \rho} $ the Weyl tensor perturbation,
 \bea
n^\alpha &=& \ft{1}{2 \Sigma} (r^2+a^2,-\Delta_r(r),0,a) \quad, \quad  \overline{m}^\alpha=\ft{1}{\sqrt{2} (r-{\rm i} a \cos\theta)}
\left(-{\rm i} a \sin\theta,0,1,-{{\rm i} \over \sin\theta} \right)
\eea
and ${\bf x}=(t,r,\theta,\phi)$. The corresponding stress energy tensor is ${\cal T}=2 \rho^{-4} T_4$,  that after being Fourier transformed produces the source term
       \begin{equation}
T (r)=-{1\over \sqrt{2\pi}} \int d\chi d\phi  dt  {}_{-2}S_{\ell m} (\chi) e^{-im\varphi+i\omega t}{}  _{-2} {\cal T}({\bf x} )
\label{eqsource}
\end{equation}
where
\begin{eqnarray}
&&_{-2} {\cal T}({\bf x} )  = 2\rho^3 L_{-1}[\rho^{-4}L_0
(\rho^{-2}{\bar \rho}^{-1}T_{nn})]   + \sqrt{2}\rho^3  \Delta^2 L_{-1}[\rho^{-4}
{\bar \rho}^2 J_+(\rho^{-2}{\bar \rho}^{-2}\Delta^{-1}
T_{{\bar m}n})]  \nn\\
&&\quad +\rho^3 \Delta^2 J_+[\rho^{-4}J_+
(\rho^{-2}{\bar \rho}T_{{\bar m}{\bar m}})] + \sqrt{2}\rho^3 \Delta^2 J_+[\rho^{-4}
{\bar \rho}^2 \Delta^{-1} L_{-1}(\rho^{-2}{\bar \rho}^{-2}
T_{{\bar m}n})] \label{eqB}\\
\nonumber
\end{eqnarray}
and
\begin{eqnarray}
 T_{n n} &=& T_{\mu \nu} n^\mu n^\nu \quad , \quad T_{n \overline{m} } =T_{\mu \nu} n^\mu \overline{m}^\nu \quad , \quad
 T_{\overline{m } \overline{m}} =T_{\mu \nu} \overline{m}^\mu \overline{m}^\nu  \nn\\
L_s&=&\partial_{\theta}+{m \over \sin\theta}
-a\omega\sin\theta+s\cot\theta \quad , \quad  L^\dagger_s=\partial_{\theta}-{m \over \sin\theta}
+a\omega\sin\theta+s\cot\theta\nn \\
J_+&=&\partial_r+{iK/\Delta}  \label{eqT}
\eea
Given the source $T_{\ell m\omega}$, the solution of the inhomogenous equation can be written in the form (\ref{rt00}-\ref{zlma0}) with the  coefficients $A_i$   given by \cite{Mino:1997bx}
\bea
A_{0}&=&{-2\rho^{-2}{\bar \rho}^{-1} \over \Delta^2}
C_{n\,n}
L_1^\dagger\{\rho^{-4}L_2^\dagger(\rho^3 S)\}
-\rho^{-3}{\bar \rho}
C_{{\bar m}\,{\bar m}}S\Bigl[
-i \partial_r \Bigl({K \over \Delta}\Bigr)-{K^2 \over \Delta^2}+
2i\rho {K \over \Delta}\Bigr]\nn\\
&& +{2\sqrt2\rho^{-3} \over \Delta}
C_{{\bar m}\,n}
\Bigl[\left(L_2^\dagger S\right)
\Bigl({iK \over \Delta}+\rho+{\bar \rho}\Bigr)
 -a\sin\theta S {K \over \Delta}({\bar \rho}-\rho)\Bigr]\nn\\
&&  ,\\
A_{1}&=&{2\sqrt2\rho^{-3}\over \Delta } C_{{\bar m}\,n}
[L_2^\dagger S+ia\sin\theta({\bar \rho}-\rho)S] -{2\rho^{-3}{\bar \rho} \over \sqrt{2\pi}}
C_{{\bar m}\,{\bar m}}S\Bigl(i{K \over \Delta}+\rho\Bigr),\nn \\
A_{2}
&=&-\rho^{-3}{\bar \rho}
C_{{\bar m}\,{\bar m}}S, \label{ais}\\
\eea
 where  $S={}_{-2}S_{\ell m}(\theta)$ and
\bea
 C_{n\,n}&=&{\mu \over 4\Sigma^3 \dot t}\left[ P_t(r^2+a^2) + a P_\phi \right]^2,\nonumber\\
C_{{\bar m}\,n}&=&
-{\mu \rho \over 2\sqrt{2}\Sigma^2 \dot t}\left[ P_t (r^2+a^2) + a P_\phi
\right]
\left[i\sin\theta\Bigl(a P_t +{P_\phi  \over \sin^2\theta}\Bigr)\right],
\label{eq:cnn}\\
C_{{\bar m}\,{\bar m}}&=&-
{\mu \rho^2 \over 2\Sigma \dot t }
\sin^2\theta
\Bigl(a P_t +{P_\phi \over \sin^2\theta}\Bigr)^2,\nonumber\\
\nonumber
\eea
\section{ Quantum period $\mathfrak{a}$ }\label{aExpansion}
Here we give (\ref{p2final2}) up to $\epsilon^4$
\bea
&&\mathfrak{a}=\left(\ell{+}\frac{1}{2}\right){{-}}\frac{(\ell (\ell{{+}}1) (15 \ell (\ell{{+}}1){+}13){+}24) \epsilon ^2}{2 \ell (\ell{+}1) (2 \ell{-}1) (2 \ell{+}1) (2 \ell{+}3)}{+}\frac{q^2  (\ell (\ell{+}1) (\ell (\ell{+}1)
   (5 \ell (\ell{+}1){-}1){+}18){+}108) m \epsilon ^3}{(\ell{-}1) \ell^2 (\ell{+}1)^2 (\ell{+}2) (2 \ell{-}1) (2 \ell{+}1) (2 \ell{+}3)}{+}\nn\\
   &&\epsilon^4\left\{\frac{m^2q^4 ({-}8160 \ell^{16}{-}65280 \ell^{15}{-}214704 \ell^{14}{-}360528 \ell^{13}{-}230118 \ell^{12}{+}335916 \ell^{11}{+}1147788 \ell^{10}{+}1846326 \ell^9}{8 (\ell{-}1) \ell^4 (\ell{+}1)^4 (\ell{+}2) (2 \ell{-}3)
   ((2 \ell{+}1)^2)^{3/2} (2 \ell{+}5) (4 \ell^2{+}4 \ell{-}3)^3}\right.\nn\\
   &&\left.{-}\frac{413946 \ell^8{+}8311080 \ell^7{+}9724656 \ell^6{-}3358854 \ell^5{-}10689684 \ell^4{-}4533264 \ell^3{+}1129248 \ell^2{+}1220832 \ell{+}233280)}{8 (\ell{-}1) \ell^4 (\ell{+}1)^4 (\ell{+}2) (2 \ell{-}3)
   ((2 \ell{+}1)^2)^{3/2} (2 \ell{+}5) (4 \ell^2{+}4 \ell{-}3)^3}\right.\nn\\
&& \left.{+}\frac{q^4 (2080 \ell^{18}{+}18720 \ell^{17}{+}80816 \ell^{16}{+}222208
   \ell^{15}{+}371610 \ell^{14}{+}197750 \ell^{13}{-}518458 \ell^{12}{-}1087194 \ell^{11}}{8 (\ell{-}1) \ell^4 (\ell{+}1)^4 (\ell{+}2) (2 \ell{-}3) ((2 \ell{+}1)^2)^{3/2} (2 \ell{+}5)(4 \ell^2{+}4
   \ell{-}3)^3}\right.\nn\\
 &&\left.{-}\frac{365930 \ell^{10}{-}1538990 \ell^9{-}1561218 \ell^8{+}2944886 \ell^7{+}6503040 \ell^6{+}3527460
   \ell^5{-}602424 \ell^4{-}1045872 \ell^3{-}233280 \ell^2)}{8 (\ell{-}1) \ell^4 (\ell{+}1)^4 (\ell{+}2) (2 \ell{-}3) ((2 \ell{+}1)^2)^{3/2} (2 \ell{+}5)(4 \ell^2{+}4
   \ell{-}3)^3}\right.\nn\\
   &&\left.{-}\frac{18480 \ell^{18}{+}166320 \ell^{17}{+}603960 \ell^{16}{+}1061760 \ell^{15}{+}613935 \ell^{14}{-}1088535 \ell^{13}{-}2775140 \ell^{12}{-}3198765
   \ell^{11}}{8 (\ell{-}1) \ell^4(\ell{+}1)^4 (\ell{+}2) (2 \ell{-}3) ((2 \ell{+}1)^2)^{3/2} (2 \ell{+}5) (4 \ell^2{+}4 \ell{-}3)^3}\right.\nn\\
   &&\left.{+}\frac{2875973 \ell^{10}{+}2753030 \ell^9{+}2614892 \ell^8{+}1950653 \ell^7{+}57648 \ell^6{-}2530951 \ell^5{-}2705934 \ell^4{-}747792 \ell^3{+}154656 \ell^2{+}51840 \ell}{8 (\ell{-}1) \ell^4
   (\ell{+}1)^4 (\ell{+}2) (2 \ell{-}3) ((2 \ell{+}1)^2)^{3/2} (2 \ell{+}5) (4 \ell^2{+}4 \ell{-}3)^3}\right\}\nn\\
\eea

\section{PN expansion of $\mathfrak{R}_{\rm in}$ }\label{mathrin}

In this appendix we display the PN expansion of  $R_{\rm in}$ and   $\mathfrak{R}_{\rm in}$  up to order $v^{10}$. The $R_n$ coefficients
up to $n=6$ are given in (\ref{rfinaln}). The remaining ones are
{\small
\bea
   &&R_7= {-}\frac{i \kappa ^7 \left(15{+}2 \mathfrak{a}\right)}{96
   \left(1{+}\mathfrak{a}\right) \left(2{+}\mathfrak{a}\right) \left(3{+}\mathfrak{a}\right) \left(1{+}2 \mathfrak{a}\right)
   \left(3{+}2 \mathfrak{a}\right)}{-}\frac{i \epsilon  \kappa ^4 \left({-}307{-}48 \mathfrak{a}{+}12
   \mathfrak{a}^2\right)}{96 \left(1{+}\mathfrak{a}\right) \left(2{+}\mathfrak{a}\right) \left(1{+}2 \mathfrak{a}\right) \left(3{+}2
   \mathfrak{a}\right)} \nn\\
   && {-}\frac{i \epsilon ^3 \left(171{-}196 \mathfrak{a}{+}80 \mathfrak{a}^2{+}16 \mathfrak{a}^3{+}16
   \mathfrak{a}^4\right)}{192 \kappa ^2 \left({-}1{+}\mathfrak{a}\right) \left({-}1{+}2 \mathfrak{a}\right)}{-}\frac{i
   \epsilon ^2 \kappa  \left({-}1375{+}2999 \mathfrak{a}{-}5334 \mathfrak{a}^2{+}3048 \mathfrak{a}^3{-}448 \mathfrak{a}^4{-}272 \mathfrak{a}^5{+}32
   \mathfrak{a}^6\right)}{64 \left({-}1{+}\mathfrak{a}\right){}^2 \left(1{+}2 \mathfrak{a}\right){}^2 \left({-}1{+}\mathfrak{a}{+}2
   \mathfrak{a}^2\right)}\nn\\
   &&R_8=\frac{\kappa ^8 \left(579{+}136 \mathfrak{a}{+}4 \mathfrak{a}^2\right)}{6144 \left(1{+}\mathfrak{a}\right)
   \left(2{+}\mathfrak{a}\right) \left(3{+}\mathfrak{a}\right) \left(4{+}\mathfrak{a}\right) \left(1{+}2 \mathfrak{a}\right)
   \left(3{+}2 \mathfrak{a}\right)}{{+}}\frac{\epsilon  \kappa ^5 \left({-}142539{-}90608 \mathfrak{a}{-}12520
   \mathfrak{a}^2{+}960 \mathfrak{a}^3{+}80 \mathfrak{a}^4\right)}{7680 \left(1{+}\mathfrak{a}\right) \left(2{+}\mathfrak{a}\right)
   \left(3{+}\mathfrak{a}\right) \left(1{+}2 \mathfrak{a}\right) \left(3{+}2 \mathfrak{a}\right) \left(5{+}2
   \mathfrak{a}\right)} \nn\\
   &&{+}\frac{\epsilon ^4 \left(315{-}216 \mathfrak{a}{-}1364 \mathfrak{a}^2{+}960 \mathfrak{a}^3{+}400 \mathfrak{a}^4{-}384
   \mathfrak{a}^5{+}64 \mathfrak{a}^6\right)}{24576 \kappa ^4 \left({-}2{+}\mathfrak{a}\right)
   \left({-}1{+}\mathfrak{a}\right)} -\frac{\epsilon ^2 \kappa ^2 \left(16538-21121 \mathfrak{a}+8858 \mathfrak{a}^2\right)}{6144
   \left(-1+\mathfrak{a}\right){}^2 \left(-1+2 \mathfrak{a}\right)} \nn\\
   && +\frac{\epsilon ^2 \kappa ^2
   \left(452961+1199243 \mathfrak{a}+1280435 \mathfrak{a}^2+989129 \mathfrak{a}^3+643034 \mathfrak{a}^4+243164 \mathfrak{a}^5+35048
   \mathfrak{a}^6+192 \mathfrak{a}^7\right)}{6144 \left(1+\mathfrak{a}\right){}^3 \left(2+\mathfrak{a}\right) \left(1+2
   \mathfrak{a}\right){}^2 \left(3+2 \mathfrak{a}\right)}\nn\\
   &&{+}\frac{\epsilon ^3 \left({-}40131{+}1823 \mathfrak{a}{+}24754 \mathfrak{a}^2{-}72276
   \mathfrak{a}^3{-}8952 \mathfrak{a}^4{+}39120 \mathfrak{a}^5{-}3744 \mathfrak{a}^6{-}1472 \mathfrak{a}^7{+}128 \mathfrak{a}^8\right)}{3072 \kappa
   \left({-}1{+}\mathfrak{a}\right){}^2 \left(1{+}2 \mathfrak{a}\right){}^2 \left(1{+}\mathfrak{a})({-}1{+}2
   \mathfrak{a}\right)}
    \nn
   \eea
   }
 {\footnotesize
   \bea
   &&R_9=\frac{i \kappa ^9 \left(19{+}2 \mathfrak{a}\right)}{1536
   \left(1{+}\mathfrak{a}\right) \left(2{+}\mathfrak{a}\right) \left(3{+}\mathfrak{a}\right) \left(4{+}\mathfrak{a}\right) \left(1{+}2
   \mathfrak{a}\right) \left(3{+}2 \mathfrak{a}\right)}{+}\frac{i \epsilon  \kappa ^6 \left({-}9449{-}4170 \mathfrak{a}{-}220
   \mathfrak{a}^2{+}40 \mathfrak{a}^3\right)}{1920 \left(1{+}\mathfrak{a}\right) \left(2{+}\mathfrak{a}\right) \left(3{+}\mathfrak{a}\right)
   \left(1{+}2 \mathfrak{a}\right) \left(3{+}2 \mathfrak{a}\right) \left(5{+}2 \mathfrak{a}\right)} \nn\\
   && {+}\frac{i \epsilon ^4
   \left(7815{-}14736 \mathfrak{a}{+}13004 \mathfrak{a}^2{-}5504 \mathfrak{a}^3{+}1104 \mathfrak{a}^4{-}256 \mathfrak{a}^5{+}64
   \mathfrak{a}^6\right)}{6144 \kappa ^3 \left({-}2{+}\mathfrak{a}\right) \left({-}1{+}\mathfrak{a}\right) \left({-}3{+}2
   \mathfrak{a}\right)}\nn\\
   && {+}\frac{i \epsilon ^3 \left({-}65151{+}72211 \mathfrak{a}{-}94782 \mathfrak{a}^2{-}98948 \mathfrak{a}^3{+}41672
   \mathfrak{a}^4{+}14992 \mathfrak{a}^5{-}10144 \mathfrak{a}^6{-}1728 \mathfrak{a}^7{+}128 \mathfrak{a}^8\right)}{768
   \left({-}1{+}\mathfrak{a}\right){}^2 \left(1{+}2 \mathfrak{a}\right){}^2 \left({-}3{+}\mathfrak{a}{+}8 \mathfrak{a}^2{+}4
   \mathfrak{a}^3\right)}\nn\\
   && {+}\frac{i \epsilon ^2 \kappa ^3 \left({-}66073{+}28523 \mathfrak{a}{-}52995 \mathfrak{a}^2{-}340815
   \mathfrak{a}^3{-}30636 \mathfrak{a}^4{+}67188 \mathfrak{a}^5{-}25616 \mathfrak{a}^6{-}14288 \mathfrak{a}^7{-}2880 \mathfrak{a}^8{+}192
   \mathfrak{a}^9\right)}{1536 \left({-}1{+}\mathfrak{a}\right){}^2 \left(1{+}\mathfrak{a}\right){}^3 \left(2{+}\mathfrak{a}\right)
   \left({-}1{+}2 \mathfrak{a}\right) \left(1{+}2 \mathfrak{a}\right){}^2 \left(3{+}2 \mathfrak{a}\right)}\nn\\
   &&R_{10}=
   {-}\frac{\kappa ^{10} \left(883{+}168 \mathfrak{a}{+}4
   \mathfrak{a}^2\right)}{122880 \left(1{+}\mathfrak{a}\right) \left(2{+}\mathfrak{a}\right) \left(3{+}\mathfrak{a}\right)
   \left(4{+}\mathfrak{a}\right) \left(5{+}\mathfrak{a}\right) \left(1{+}2 \mathfrak{a}\right) \left(3{+}2
   \mathfrak{a}\right)} \nn\\
   &&{+}\frac{\epsilon  \kappa ^7 \left(24325247{+}19846250 \mathfrak{a}{+}5174568
   \mathfrak{a}^2{+}374640 \mathfrak{a}^3{-}19600 \mathfrak{a}^4{-}1120 \mathfrak{a}^5\right)}{860160 \left(1{+}\mathfrak{a}\right)
   \left(2{+}\mathfrak{a}\right) \left(3{+}\mathfrak{a}\right) \left(4{+}\mathfrak{a}\right) \left(1{+}2 \mathfrak{a}\right)
   \left(3{+}2 \mathfrak{a}\right) \left(5{+}2 \mathfrak{a}\right) \left(7{+}2 \mathfrak{a}\right)}   \nn\\
   &&{+}\frac{\epsilon ^5
   \left(2835{-}2574 \mathfrak{a}{-}11844 \mathfrak{a}^2{+}11368 \mathfrak{a}^3{+}1680 \mathfrak{a}^4{-}4256 \mathfrak{a}^5{+}1344 \mathfrak{a}^6{-}128
   \mathfrak{a}^7\right)}{491520 \kappa ^5 \left({-}2{+}\mathfrak{a}\right)
   \left({-}1{+}\mathfrak{a}\right)} + \frac{\epsilon ^3 \kappa  \left(80080-88187 \mathfrak{a}+32182 \mathfrak{a}^2\right)}{24576
   \left(-1+\mathfrak{a}\right){}^2 \left(-1+2 \mathfrak{a}\right)} \nn\\
   && -\frac{\epsilon ^3 \kappa
   \left(2222991+6279517 \mathfrak{a}+6868237 \mathfrak{a}^2+4595739 \mathfrak{a}^3+2348214 \mathfrak{a}^4+765780
   \mathfrak{a}^5+98232 \mathfrak{a}^6-1984 \mathfrak{a}^7+128 \mathfrak{a}^8\right)}{24576 \left(1+\mathfrak{a}\right){}^3
   \left(2+\mathfrak{a}\right) \left(1+2 \mathfrak{a}\right){}^2 \left(3+2 \mathfrak{a}\right)}\nn\\
   && {+}\frac{\epsilon ^4 \left(589461{-}1477833 \mathfrak{a}{+}1347936 \mathfrak{a}^2{+}403184
   \mathfrak{a}^3{-}1641824 \mathfrak{a}^4{+}1097248 \mathfrak{a}^5{-}232448 \mathfrak{a}^6{-}13568 \mathfrak{a}^7{+}6400 \mathfrak{a}^8{-}256
   \mathfrak{a}^9\right)}{98304 \kappa ^2 \left({-}2{+}\mathfrak{a}\right) \left({-}1{+}\mathfrak{a}\right){}^2
   \left(1{+}\mathfrak{a}\right) \left({-}3{+}2 \mathfrak{a}\right) \left({-}1{+}2 \mathfrak{a}\right)} \nn\\
 &&+   \frac{\epsilon ^2 \kappa ^4 \left(133673+213680 \mathfrak{a}-1060336 \mathfrak{a}^2+4791350 \mathfrak{a}^3-6624937
   \mathfrak{a}^4+2833370 \mathfrak{a}^5\right)}{491520 \left(-1+\mathfrak{a}\right){}^2 \left(1+\mathfrak{a}\right){}^3
   \left(-1+2 \mathfrak{a}\right)} \nn\\
   && -\frac{\epsilon ^2 \kappa ^4 \left(99714618+1389963
   \mathfrak{a}+241124572 \mathfrak{a}^2+366858848 \mathfrak{a}^3+162319984 \mathfrak{a}^4+22669520 \mathfrak{a}^5\right)}{491520
   \left(2+\mathfrak{a}\right) \left(3+\mathfrak{a}\right) \left(1+2 \mathfrak{a}\right){}^2 \left(3+2 \mathfrak{a}\right)
   \left(5+2 \mathfrak{a}\right)}
\eea
}
The  $\mathfrak{R}_n$ coefficients up to $n=6$ are given in (\ref{gothicR}). The remaining ones are

\bea
&&\mathfrak{R}_7={-}\frac{i \left(12{-}18 \ell{+}13 \ell^2{+}12 \ell^3{+}5
	\ell^4\right) \epsilon ^3}{24 \ell ({-}1+2 \ell) \kappa ^2}{-}\frac{i \left(-14+39 \ell+9 \ell^2\right) \epsilon  \kappa ^4}{48 (1{+}\ell) (2{+}\ell) (3{+}2 \ell) (5{+}2 \ell)}{-}\frac{i (8{+}\ell) \kappa ^7}{24 (1{+}\ell)
	(2{+}\ell) (3{+}2 \ell) (5{+}2 \ell) (7+2 \ell)}\nn\\
&&-\frac{i \left(-108+1428 \ell+2983 \ell^2+2853 \ell^3+1561 \ell^4+431 \ell^5+152 \ell^6+84 \ell^7+16 \ell^8\right) \epsilon ^2 \kappa }{8 \ell
	(1+\ell)^3 (3+2 \ell)^2 \left(-1+4 \ell^2\right)}\nn\\
&&\mathfrak{R}_8=\frac{\left(3456+13572 \ell+24540 \ell^2+30720 \ell^3+27019 \ell^4+14641 \ell^5+4489 \ell^6+835 \ell^7+124 \ell^8+4 \ell^9\right) \epsilon ^3}{48 \ell (1+\ell)^2 (3+2 \ell)^2 \left(-1+4 \ell^2\right)
	\kappa }\nn\\
&&{+}\frac{\left({-}133920+72972 \ell{+}788796 \ell^2{+}1157009 \ell^3{+}746761 \ell^4{+}138424 \ell^5{-}83738 \ell^6{-}32675 \ell^7{+}9769 \ell^8\right) \epsilon ^2
	\kappa ^2}{96 \ell (1{+}\ell)^3 (2{+}\ell) (-1{+}2 \ell) (1{+}2 \ell) (3{+}2 \ell)^3 (5{+}2 \ell)}\nn\\
&&+\frac{(6758 \ell^9{+}1020 \ell^{10}{+}24 \ell^{11}) \epsilon ^2\kappa^2}{96 \ell (1{+}\ell)^3 (2{+}\ell) (-1{+}2 \ell) (1{+}2 \ell) (3{+}2 \ell)^3 (5{+}2 \ell)}+\frac{\left({-}4368{-}458 \ell{+}815 \ell^2{+}190 \ell^3{+}5 \ell^4\right) \epsilon  \kappa ^5}{480 (1{+}\ell) (2{+}\ell) (3{+}\ell)
	(3{+}2 \ell) (5{+}2 \ell) (7{+}2 \ell)}\nn\\
&&{+}\frac{\left(162{+}35 \ell{+}\ell^2\right) \kappa ^8}{384 (1{+}\ell) (2{+}\ell) (3{+}2 \ell) (5{+}2 \ell) (7{+}2 \ell) (9{+}2 \ell)}+\frac{\ell \left(-12+4 \ell+15 \ell^2-5 \ell^3-3 \ell^4+\ell^5\right) \epsilon ^4}{96 \left(3-8 \ell+4 \ell^2\right) \kappa
	^4}
	\eea
{\small	
	\bea
&&\mathfrak{R}_9=\frac{i \left(4104{+}49248
	\ell{+}129288 \ell^2{+}178341 \ell^3{+}151139 \ell^4{+}75610 \ell^5{+}18766 \ell^6{+}1093 \ell^7{-}53 \ell^8{+}124 \ell^9{+}20 \ell^{10}\right) \epsilon ^3}{48 \ell (1+\ell)^3 (2+\ell) (-1+2 \ell) (1+2 \ell) (3+2 \ell)^2}\nn\\
&&+\frac{i
	\left(48-96 \ell+104 \ell^2-44 \ell^3+\ell^4-4 \ell^5+3 \ell^6\right) \epsilon ^4}{48 (-1+\ell) (-3+2 \ell) (-1+2 \ell) \kappa ^3}+\frac{i (10+\ell) \kappa ^9}{192 (1+\ell) (2+\ell)
	(3+2 \ell) (5+2 \ell) (7+2 \ell) (9+2 \ell)}\nn\\
&&+\frac{i \left({-}176400{-}128328 \ell{+}431016 \ell^2{+}760226 \ell^3{+}394817
	\ell^4{-}135439 \ell^5{-}273319 \ell^6{-}136603 \ell^7{-}27238 \ell^8\right) \epsilon ^2 \kappa ^3}{96 \ell (1+\ell)^3 (2+\ell)^2 (-1+2 \ell) (1+2 \ell) (3+2 \ell)^3 (5+2
	\ell)}\nn\\
&&+\frac{i\left({+}132 \ell^9{+}840 \ell^{10}{+}96 \ell^{11}\right) \epsilon ^2 \kappa ^3}{96 \ell (1+\ell)^3 (2+\ell)^2 (-1+2 \ell) (1+2 \ell) (3+2 \ell)^3 (5+2
	\ell)}+\frac{i \left(-1426-195 \ell+110 \ell^2+15 \ell^3\right) \epsilon  \kappa ^6}{480 (1{+}\ell) (2{+}\ell) (3{+}\ell) (3{+}2 \ell) (5{+}2 \ell) (7{+}2 \ell)}\nn\\
&&\mathfrak{R}_{10}=-\frac{  \ell (\ell^2{-}1) (\ell^2{-}4) (\ell{-}3)(\ell{-}4) \epsilon ^5}{960 (2\ell{-}1)(2\ell{-}3) \kappa^5}
{-}\frac{\left(33696{-}44064 \ell{-}42336 \ell^2{+}29856 \ell^3\right)
		\epsilon ^4}{192 (-1{+}\ell) \ell (1{+}\ell) ({-}3{+}2 \ell) ({-}1{+}2 \ell)^3 (1{+}2 \ell) (3{+}2 \ell)^2 \kappa ^2}\nn\\
&&-\frac{\left(18375 \ell^4+53291 \ell^5+58594 \ell^6-89074 \ell^7-82437 \ell^8+23703 \ell^9+23580 \ell^{10}+1792 \ell^{11}+608 \ell^{12}+16 \ell^{13}\right)
	\epsilon ^4}{192 (-1+\ell) \ell (1+\ell) (-3+2 \ell) (-1+2 \ell)^3 (1+2 \ell) (3+2 \ell)^2 \kappa ^2}\nn\\
&&-\frac{\left(-302400+1201680 \ell+5744016 \ell^2+9652988 \ell^3+9102146 \ell^4+5067989
	\ell^5+1465614 \ell^6+71054 \ell^7\right) \epsilon ^3 \kappa }{192 \ell (1+\ell)^3 (2+\ell) (-1+2 \ell) (1+2 \ell) (3+2 \ell)^3 (5+2
	\ell)}\nn
\eea
\bea
&&{-}\frac{\left({-}68374 \ell^8{-}9267 \ell^9{+}2454 \ell^{10}{+}492 \ell^{11}{+}8 \ell^{12}\right) \epsilon ^3 \kappa }{192 \ell (1{+}\ell)^3 (2{+}\ell) (-1{+}2 \ell) (1{+}2 \ell) (3{+}2 \ell)^3 (5{+}2
	\ell)}{+}\frac{(1{+}\ell)^{-3}\left(71517600{+}139291920 \ell{+}6230364 \ell^2\right) \epsilon ^2 \kappa ^4}{960 \ell  (2{+}\ell)^2 (3{+}\ell) (-1{+}2 \ell) (1{+}2 \ell) (3{+}2 \ell)^3 (5{+}2 \ell)^2 (7{+}2
	\ell)}\nn\\
&&{+}\frac{\left({-}196339512 \ell^3{-}169789499 \ell^4{+}50239798 \ell^5{+}192489961 \ell^6{+}163418256 \ell^7{+}74755739 \ell^8{+}19776642 \ell^9\right) \epsilon ^2 \kappa ^4}{960 \ell (1{+}\ell)^3 (2{+}\ell)^2 (3{+}\ell) (-1{+}2 \ell) (1{+}2 \ell) (3{+}2 \ell)^3 (5{+}2 \ell)^2 (7{+}2
	\ell)}\nn\\
&&+\frac{\left(2594195 \ell^{10}-17824
	\ell^{11}-51880 \ell^{12}-5280 \ell^{13}-80 \ell^{14}\right) \epsilon ^2 \kappa ^4}{960 \ell (1+\ell)^3 (2+\ell)^2 (3+\ell) (-1+2 \ell) (1+2 \ell) (3+2 \ell)^3 (5+2 \ell)^2 (7+2
	\ell)}\nn\\
&&+\frac{\left(628272+340124 \ell+26544 \ell^2-11445 \ell^3-1820 \ell^4-35 \ell^5\right) \epsilon  \kappa ^7}{26880 (1+\ell) (2+\ell) (3+\ell) (4+\ell) (3+2 \ell) (5+2 \ell) (7+2 \ell) (9+2
	\ell)}\nn\\
&&-\frac{\left(242+43 \ell+\ell^2\right) \kappa ^{10}}{3840 (1+\ell) (2+\ell) (3+2 \ell) (5+2 \ell) (7+2 \ell) (9+2 \ell) (11+2 \ell)}\nn\\
\eea
}

 Finally plugging (\ref{epsilonkappa}) into (\ref{gothicR0}) and using (\ref{p2final}), (\ref{gothicR}) one finds (up to $v^4$)
 \bea
 \mathfrak{R}_{\rm in} (r)     &=&\frac{2 M \Gamma (\ell-1) (-2 i r \omega )^{\ell+3} r^{\ell+2} }{\Gamma (2\ell +2)}  \left[  1+\frac{2 i r  \omega }{1+\ell} -\frac{(2+\ell) M}{r}-\frac{(9+\ell) r^2 \omega ^2}{6+10 \ell+4 \ell^2}  \right. \label{rrl} \\
&&  -\frac{i (4+\ell) r^3 \omega^3}{(1+\ell) \left(6+7 \ell+2 \ell^2\right)}
+ M \omega  \left(2 i \log \left(4 M  \omega \right)+\pi -2 i \psi (\ell-1)-\frac{i (3 \ell+7)}{\ell+1} \right) \nn\\
&& +\frac{(-1+\ell) (1+\ell) (2+\ell) M^2}{(-1+2 \ell) r^2}+\frac{\left(50+19 \ell+\ell^2\right) r^4 \omega ^4}{8 (1+\ell) (2+\ell) (3+2\ell) (5+2 \ell)} \nn\\
&& \left. +M r \omega ^2
   \left(\frac{(4+\ell) (9+\ell)}{2 (1+\ell) (3+2 \ell)}+\frac{2 i \pi -4 \log \left(4 M v^3 \omega \right)+4 \psi _0(\ell-1)}{\ell+1}\right)
+\ldots   \right]  \nn
\label{gothicrin}\eea

\end{appendix}

\providecommand{\href}[2]{#2}\begingroup\raggedright\endgroup

\end{document}